\documentclass{aa}  

\usepackage{graphicx}
\usepackage{txfonts}
\usepackage{amsmath}	
\usepackage{amssymb}	
\usepackage{multicol}   
\usepackage{pdflscape}	
\usepackage{float}
\usepackage[export]{adjustbox}

\usepackage{wasysym}   
\usepackage{enumitem}  
\setlist[enumerate]{label={\arabic*.}}

\usepackage[breaklinks=true, colorlinks=true, citecolor=blue, linkcolor=blue,urlcolor=blue]{hyperref}

\newcommand{\cloudy}{{\footnotesize CLOUDY}\xspace}
\newcommand{\beagle}{{\footnotesize BEAGLE}\xspace}

\newcommand{\bpass}{{\footnotesize BPASS}\xspace}
\newcommand{\sevn}{{\footnotesize SEVN}\xspace}
\newcommand{\parsec}{{\footnotesize PARSEC}\xspace}
\newcommand{\mesa}{{\footnotesize MESA}\xspace}
\newcommand{\galaxev}{{\footnotesize GALAXEV}\xspace}

\newcommand{\HST}{\textit{HST}\xspace}
\newcommand{\JWST}{\textit{JWST}\xspace}
\newcommand{\GLS}{{\footnotesize GALSEVN}\xspace}

\newcommand{\Msun}{\hbox{M$_{\rm{\odot}}$}\xspace}

\newcommand{\Lsun}{\hbox{L$_{\rm{\odot}}$}\xspace}

\newcommand{\tprime}{\hbox{$t^\prime$}\xspace}
\newcommand{\hii}{\hbox{H\,{\sc ii}}\xspace}

\newcommand{\zism}{\hbox{$Z_\mathrm{ISM}$}\xspace}

\newcommand{\nh}{\hbox{$n_{\mathrm{H}}$}\xspace}

\newcommand{\xid}{\hbox{$\xi_{\rm{d}}$}\xspace}
\newcommand{\fesc}{\hbox{$f_{\rm{esc}}$}\xspace}

\newcommand{\CO}{\hbox{C/O}\xspace}

\newcommand{\Uav}{\hbox{$\langle U\rangle$}\xspace}

\newcommand{\logU}{\hbox{$\log \Uav$}\xspace}

\newcommand{\mchar}{\hbox{$m_{\mathrm{char}}$}\xspace}

\newcommand{\xiion}{\hbox{$\xi_{\rm{ion}}$}\xspace}
\newcommand{\xiionst}{\hbox{$\xi_{\rm{ion}}^{*}$}\xspace}
\newcommand{\xiionHII}{\hbox{$\xi_{\rm{ion}}^{\rm{HII}}$}\xspace}

\newcommand{\lya}{\hbox{Ly$\alpha$}\xspace}

\newcommand{\heii}{\hbox{He\,\textsc{ii}\,$\lambda1640$}\xspace}

\newcommand{\heiiopt}{\hbox{He\,\textsc{ii}\,$\lambda4686$}\xspace}
\newcommand{\hb}{\hbox{H$\beta$}\xspace}

\newcommand{\ha}{\hbox{H$\alpha$}\xspace}

\newcommand{\lhei}{\hbox{He\,\textsc{i}}\xspace}

\newcommand{\lheii}{\hbox{He\,\textsc{ii}}\xspace}

\usepackage{xspace}
\usepackage[dvipsnames]{xcolor}
\usepackage{orcidlink}
\usepackage[english]{babel}
\usepackage{breqn}

\begin{document} 

\title{A new prescription for the spectral properties of population III stellar populations}

\author{
Marie~Lecroq\inst{1}\orcidlink{0009-0008-2198-9651}
\and Stéphane~Charlot\inst{1}\orcidlink{0000-0003-3458-2275}
\and Alessandro~Bressan\inst{2,3}\orcidlink{0000-0002-7922-8440}
\and Gustavo~Bruzual\inst{4}\orcidlink{0000-0002-6971-5755}
\and Guglielmo~Costa\inst{5}\orcidlink{0000-0002-6213-6988}
\and Giuliano~Iorio\inst{6,3,7}\orcidlink{0000-0003-0293-503X}
\and Michela~Mapelli\inst{8,6,7}\orcidlink{0000-0001-8799-2548}
\and Filippo~Santoliquido\inst{9,10}\orcidlink{0000-0003-3752-1400}
\and Kendall~Shepherd\inst{2,3}\orcidlink{0000-0001-5231-0631}
\and Mario~Spera\inst{2}\orcidlink{0000-0003-0930-6930}
}

\institute{
Sorbonne Universit\'e, CNRS, UMR 7095, Institut d'Astrophysique de Paris, 98 bis bd Arago, 75014 Paris, France\\
\email{lecroq@iap.fr; charlot@iap.fr}
\and 
SISSA, via Bonomea 265, I-34136 Trieste, Italy
\and 
INAF, Osservatorio Astronomico di Padova, Vicolo dell’Osservatorio 5, I–35122, Padova, Italy
\and 
Instituto de Radioastronom{\'i}a y Astrof{\'i}sica, UNAM, Campus Morelia, Michoacan, M{\'e}xico, C.P. 58089, M{\'e}xico
\and 
Univ Lyon1, ENS de Lyon, CNRS, Centre de Recherche Astrophysique de Lyon UMR5574, F-69230 Saint-Genis-Laval, France
\and 
Dipartimento di Fisica e Astronomia Galileo Galilei, Università di Padova, Vicolo dell’Osservatorio 3, I–35122 Padova, Italy
\and 
INFN-Padova, Via Marzolo 8, I–35131 Padova, Italy\
\and 
Institut für Theoretische Astrophysik, ZAH, Universität Heidelberg, Albert-Überle-Straße 2, D-69120, Heidelberg, Germany
\and 
Gran Sasso Science Institute (GSSI), Viale Francesco Crispi 7, 67100, L’Aquila, Italy
\and 
INFN, Laboratori Nazionali del Gran Sasso, I-67100 Assergi, Italy
}

\date{Received 2 October 2024 / Accepted 19 January 2025}

 
\abstract{
We investigated various emission properties of extremely low metallicity stellar populations in the Epoch of Reionization (EoR), using the new \GLS model, which has shown promising agreement between spectral predictions and observations at lower redshifts and higher metallicities. We find that emission-line diagnostics previously proposed to discriminate between population III (Pop\,III) stars and other primordial ionizing sources are effective, but only for stellar-population ages below $\sim1$\,Myr. We provide other key quantities relevant to modeling Pop\,III stellar populations in the EoR, such as the production efficiency of ionizing photons, which is critical for reionization studies, the production rate of Lyman-Werner photons, which can dissociate H$_2$ and influence the efficiency of star formation, and the rates of different types of supernov\ae, offering insights into the timescales of chemical enrichment in metal-poor environments. We complement our study with a self-consistent investigation of the gravitational-wave signals generated by the mergers of binary black holes that formed through stellar evolution and their detectability. The results presented here provide valuable predictions for the study of the EoR, on the crucial role of low-metallicity stellar populations in reionization mechanisms and star formation, as well as meaningful insights into potential observational counterparts to direct detections of Pop\,III stars.
}


\maketitle
\section{Introduction}
\label{sec:intro}

One of the most fundamental yet still largely unexplored phases in the history of our Universe is the Epoch of Reionization (EoR), corresponding to the complex phase transition from an essentially neutral to an almost fully ionized state for intergalactic gas in the redshift range $5\la z\la20$ \citep[e.g.,][]{dayal2018, bosman2022, robertson2022, padmanabhan2024}. The approximate scenario of reionization, based on the idea that the first light-emitting objects ionize their immediate surroundings and create expanding ionized bubbles, is straightforward in principle, but still suffers from significant simplifications and unresolved issues. One of the main uncertainties lies in determining the mechanisms controlling the formation of the first luminous objects. 

The main sources of ionizing photons in the EoR are thought to be the first generations of massive stars in nascent galaxies \citep[e.g.,][]{madau1999, bouwens2015, robertson2022}. The very first stars to form presumably were massive \citep[up to $\sim1000\,\Msun$; e.g.,][]{schauer2020}, nearly metal-free, exotic population\,III (hereafter simply Pop\,III) stars. Their very short lifetimes and the associated short timescales of chemical enrichment meant that more metal-rich, population\,II (Pop\,II) stars probably quickly took control of the emission in most primeval galaxies \citep[e.g.,][]{bromm2004}.
Yet, recent studies have shown that Pop\,III star formation could have persisted down to redshifts $z \sim\,6$ \citep[e.g.,][]{mebane2018, hartwig2022, venditti2023}.
Other sources are likely to have contributed significantly to the reionization of the Universe, in particular, active galactic nuclei \citep[AGNs; e.g.,][]{wang2010,parsa2018, harikane2023, maiolino2023b} and accretion disks of putative, direct-collapse black holes \citep[DHCBs; e.g.,][]{begelman2006a, inayoshi2020}.
Understanding the radiative properties of young, extremely metal-poor stellar populations in their pristine environment is therefore key to constraining the role played by primeval galaxies as drivers of reionization.

Recent activity in this field has been boosted by the advent of the {\it James Webb Space Telescope} (\JWST), which has collected spectra of galaxies out to redshifts $z\sim14$, well into the EoR \citep[e.g.,][]{curtislake2023, fujimoto2023, carniani2024, robertsborsani2024}, as well as by growing efforts to understand the influence of the first stellar generations on the timing and depth of the cosmological 21-cm signal \citep[e.g.,][]{mirocha2018, mebane2020, gesseyjones2022, ventura2023, pochinda2024}. One of the main objectives of \JWST observations is to detect direct signatures of Pop\,III stars, in particular through the transient H- and \lheii-line emission of the pristine gas from which they form and photoionize, before it is enriched by metals from the first supernova (SN) explosions \citep[e.g.,][]{tumlinson2000, schaerer2003, inoue2011}. The challenge of such detections lies both in the faintness of the targets, which can be mitigated by strong gravitational lensing in cluster-caustic transits \citep[e.g.,][]{zackrisson2023}, and in verifying the nature of Pop\,III stars, because of limited line diagnostics and the variety of alternative sources such as Pop\,II stars, AGNs and DCBHs \citep[e.g.,][]{venditti2023}. While several Pop\,III star candidates have been identified through the tentative detection of \lheii emission \citep[][]{wang2022, maiolino2023a, vanzella2023}, their confirmation is still pending.

Some of the currently widely used predictions for the spectral properties of Pop\,III stellar populations are those presented by \citet[][see also \citealt{raiter2010}]{schaerer2002, schaerer2003}, based on the spectral-synthesis code of \citet{schaerer1998}. These predictions were among the first \citep[along with those of][]{tumlinson2000, bromm2001} to incorporate zero-metallicity stars, and have remained the benchmark for Pop\,III modeling ever since they were published (although we note the adoption by \citealt{gesseyjones2022} of the more recent stellar-evolution code \mesa by \citealt{paxton2019}). They have notably been used by \citet{nakajima2022} to derive observational criteria to identify spectral features characteristic of extremely young Pop\,III stars. Yet, a potential limitation of the \citet{schaerer2003} predictions is that they are based on single-star population models, while observations of \lheii emission in metal-poor (Pop\,II), actively star-forming galaxies indicate the importance of including the hard ionizing radiation from binary-star processes \citep[e.g.,][]{eldridge2012, berg2018}. In fact, even the popular \bpass model of \citet{stanway2018}, which includes such processes, struggles to reproduce the strong \lheii emission observed in some low-metallicity starburst galaxies \citep{stanway2019}. The recent predictions by \citet{lecroq2024}, based on the new \GLS binary-star population synthesis model, provide notably better agreement with observations at metallicities probed down to about 3\% of solar. 

In this paper, we  present a novel approach to modeling the spectral properties of extremely metal-poor stars in early universe studies, using the new \GLS spectral-synthesis model introduced by \citet{lecroq2024} to carry out an in-depth study of the emission properties of Pop\,III ($Z=10^{-11}$) and low-metallicity Pop\,II ($10^{-6}\leq Z\leq 10^{-3}$) stellar populations.\footnote{The Pop\,III stellar metallicity $Z=10^{-11}$ quoted in this work refers to the mass fraction of all elements heavier than lithium in the stellar-evolution calculations of \citet{costa2023}. The actual stellar metallicity of these models, including primordial Li, is $Z\approx2.7\times10^{-9}$  \citep{costa2025}. The value $Z=10^{-11}$ was selected because the features specific to Pop\,III star evolution (such as the critical role of the triple-alpha reaction to activate the CNO cycle; see Sect.~\ref{sec:popIII_res_diags}) appear at $Z\leq10^{-10}$ \citep[e.g.,][]{cassisi1993, marigo2001}. For Pop\,II stars, the quoted metallicities ($10^{-6}\leq Z\leq 10^{-3}$) refer in the standard way to the mass fraction of all elements heavier than helium.} Specifically, we start by examining the predictions of these models for the H-, \lhei-, and \lheii-ionizing photon rates and the emission-line diagnostics proposed by \citet{nakajima2022} to distinguish Pop\,III stellar emission from that of other potential ionizing sources in the early Universe. We also investigate other emission properties of pristine stellar populations critical to EoR studies: the production efficiency of ionizing photons, that is, the ratio \xiion of the H-ionizing photon rate to the nonionizing ultraviolet (UV) luminosity, which makes it possible to link the UV luminosity function of galaxies to reionization \citep[e.g.,][]{robertson2022}; and the production rate of Lyman-Werner (LW) photons (with energies in the range 912--1150\,\AA) capable of dissociating molecular hydrogen, which control the efficiency of star formation in the early universe \citep[e.g.,][]{haiman2000, oh2002, agarwal2012, incatasciato2023}. Another advantage of the \GLS model is that it allows us to compute, in a self-consistent way with light production, stellar-population properties related to end-of-life events and the evolution of binary remnants, such as: the rates of different types of SNe, which control the onset of chemical enrichment associated with these Pop\,III stellar populations \citep[e.g.,][]{heger2010, goswami2022, vanni2023}; and the gravitational-wave (GW) signal associated with the merging over time of the binary black hole (BBH) remnants they produce \citep[e.g.,][]{kinugawa2014,kinugawa2016, hartwig2016, liu2020,tanikawa2021, tanikawa2022,tanikawa2024,wang2022,santoliquido2023,liu2024,mestichelli2024}. Such self-consistent predictions should be particularly useful for the coherent modeling of the multiple observational signatures of Pop\,III stellar populations in the framework of galaxy formation theories.

The paper layout is as follows: we describe our approach to model the stellar and nebular emission from Pop\,III and and low-metallicity Pop\,II stellar populations using the \GLS model in Sect.~\ref{sec:popIII_reion}. In Sect.~\ref{sec:popIII_res}, we present the various model properties mentioned above and examine the location of the models in the emission-line diagnostic diagrams proposed by \citet{nakajima2022} to characterize ionizing sources in the early Universe. We investigate the gravitational-wave signatures of the merging over time of BBHs issued from these stellar populations in Sect.~\ref{sec:popIII_gw}. Section~\ref{sec:popIII_dsc} summarizes our results.

\section{Spectral modeling}
\label{sec:popIII_reion}

In this section, we describe the observational signatures of reionization era galaxies we focus on reproducing, and the models used to compute them. 

As described by \citet{lecroq2024}, we followed the approach of \citet[][see also \citealt{gutkin2016}]{charlot2001} and expressed the luminosity per unit frequency $\nu$ emitted at time $t$ by a star-forming galaxy as
\begin{equation}
L_{\nu}(t)=\int_0^t \mathrm{d}\tprime\, \psi(t-\tprime) \, S_{\nu}[\tprime,Z(t-\tprime)] \, T_{\nu}(t,\tprime)\,,
\label{eq:flux_gal}
\end{equation}
where $\psi(t-\tprime)$ is the star formation rate at time $t-\tprime$, $S_\nu[\tprime,$ $Z(t-\tprime)]$ is the luminosity produced per unit frequency and per unit mass by a single generation of stars of age $\tprime$ and metallicity $Z(t-\tprime)$, and $T_\nu(t,\tprime)$ is the transmission function of the interstellar medium (ISM). The modeling procedure followed in this work is similar to that presented by \citet{lecroq2024}, with $S_\nu$ computed using the code \protect\GLS for populations of single and binary stars, combined with the calculation of $T_\nu$ using the \cloudy\ photoionization code \citep{ferland2017}.

\subsection{Stellar population modeling}
\label{sec:popIII_reion_GLS}

\protect\GLS models the emission from populations of single and binary stars by combining the population synthesis code \sevn \citep{spera2015,spera2017,spera2019,mapelli2020,iorio2023} with the spectral evolution predictions from \galaxev \citep{bruzual2003}.

\sevn (Stellar EVolution for N-body)\footnote{\url{https://gitlab.com/sevncodes/sevn}. The \sevn version used in this work is the release {\it Iorio22} (\url{https://gitlab.com/sevncodes/sevn/-/releases/iorio22}).} relies on the interpolation of stellar properties from evolutionary track libraries, while accounting for binary-evolution processes by allowing jumps between stellar tracks. In this work, we adopted the same libraries as presented by \citet{lecroq2024}, namely the \parsec \citep[PAdova and TRieste Stellar Evolution Code; ][]{bressan2012,chen2015,costa2019,costa2023,nguyen2022,volpato2023} evolutionary tracks for nonrotating stars with initial masses between 2 and 600 \Msun and metallicity as low as 10$^{-11}$. 
Details on the prescriptions used for Pop\,III stars can be found in \citet{volpato2023}. 
The prescriptions for supernova (SN) remnants and natal kicks are the same as those described in section~2.1 of \citet{lecroq2024}. They account for electron-capture \citep[][]{giacobbo2019}, core-collapse \citep{fryer2012}, and pair-instability \citep{mapelli2020} SNe, relying on a delayed-SN model to predict a smooth transition between maximum neutron-star mass and minimum black-hole mass. Natal kicks are generated in agreement with the proper-motion distribution of young Galactic pulsars \citep{hobbs2005}, with reduced kick magnitudes for stripped and ultra-stripped SNe \citep{tauris2017}, as described by \cite{giacobbo2020}.

An in-depth description of \sevn handling of binary-evolution processes and products can be found in section~2.3 of \citet{iorio2023}, as well as in appendix~A of \citet{lecroq2024}. These processes notably include mass transfer, driven either by winds or Roche-lobe overflow, common-envelope evolution, quasi-homogeneous evolution \citep[QHE; ][]{eldridge2011, iorio2023},\footnote{A star with metallicity lower than 0.004 can be spun up by the accretion of substantial material through stable Roche-lobe overflow mass transfer, leading to the replenishment of its core with fresh hydrogen via rotational mixing. The star remains fully mixed until it burns all its hydrogen into helium (ending as a pure-He star) at nearly constant radius. During this QHE phase, it brightens and its temperature increases, thereby augmenting the population of compact, hot luminous stars.} stellar mergers, the effects of magnetic braking on stellar angular-momentum, changes in angular momentum and orbital motion due to stellar tides, and orbital decay due to the emission of gravitational waves \citep{hurley2002}. As in the aforementioned works, we assumed in this paper a stable mass transfer for donor stars which are on the main sequence and in the Hertzprung gap, whereas mass transfer stability depends on the binary mass ratios and physical properties of donors in later evolution stages.
Finally, the initial conditions we adopted in this work are the same as presented by \citet{lecroq2024}. In brief, we produced a stochastic population of $10^6$ evolving binary pairs, with primary-star masses in the range $2\leq m\leq300\,\Msun$ drawn from the initial mass function (IMF) of interest. The ratio of initial secondary-star mass to initial primary-star mass ($q$), the orbital period ($P$) and the eccentricity ($e$) were then drawn from the corresponding probability density functions (PDFs) taken from \citet{sana2012}: $\mathrm{PDF}(q) \propto q^{-0.1}$ with $q\in[0.1, 1.0]$; $\mathrm{PDF}(\mathcal{P}) \propto \mathcal{P}^{-0.55}$ with $\mathcal{P}=\log(P/\mathrm{day})\in[0.15,5.5]$;  and  $\mathrm{PDF}(e) \propto e^{-0.42}$ with $e\in[0,0.9]$.

\protect\GLS then computes the spectral evolution of \sevn predicted stellar populations by using the approach implemented in \galaxev, that is, each \sevn star is assigned a spectrum from a broad range of spectral libraries based on its evolutionary stage and its physical properties (mostly its metallicity Z and its effective temperature). The spectral libraries included in \galaxev are listed in appendix~A of \citet{sanchez2022} as well as in section~2.1 of \citet{lecroq2024}.

We also included, in the spectral energy distribution (SED) of the stellar population, the contribution from accretion disks of X-ray binaries (XRBs), computed self-consistently as detailed in appendix~B of \citet{lecroq2024}. We however did not consider the emission from fast radiative shocks from stellar winds and SNe, as these were shown to have a very small impact on the photon production rates and emission-line luminosities of interest to us in the present study \citep[see section~4.4 of][]{lecroq2024}.


\subsection{Photoionization calculation}
\label{sec:popIII_reion_photoion}

The radiative transfer of these modeled spectra through the surrounding gas-rich ISM, expressed in Eq.~\ref{eq:flux_gal} by the transmission function $T_{\nu}(t,t')$, was computed using version C17.00 of the photoionization code \cloudy \citep{ferland2017}, in the same manner as described in section~2.2 of \citet[][following \citealt{charlot2001}]{lecroq2024}, under different assumptions about the physical conditions in this gas.
We considered that galaxies are ionization-bounded and that the radiative properties of their ionized gas can be described by a set of effective parameters, the main ones being hydrogen density \nh, gas-phase metallicity \zism, carbon-to-oxygen abundance ratio \CO, dust-to-metal mass ratio \xid, and zero-age volume-averaged ionization parameter $\langle U\rangle$ \citep[e.g.,][]{gutkin2016,plat2019}. We assumed spherical geometry, with a default \hii-region inner radius of 0.1\,pc, and stopped the photoionization calculations at the radius where the electron density falls below 1\% of \nh. Furthermore, we considered the metallicity of the photoionized gas to be the same as that of the ionizing stars, consistent with the fact that we are primarily interested in emission properties up to the appearance of the first SNe. We adopted the prescription by \citet{gutkin2016} for the abundances, abundance scaling, and depletions of interstellar elements for nonzero metallicity models. More details about the choice of population synthesis related and nebular parameters for the different models presented in this work are given in the next section.

\subsection{Choice of \protect\GLS adjustable parameters}
\label{sec:popIII_reion_params}

The \protect\GLS models considered for Pop\,III stars in this work are, unless otherwise specified, pure-binary models (i.e., with unit binary fraction) at the lowest metallicity available in \protect\GLS, that is, $Z = 10^{-11}$. We also computed, for comparison, slightly more enriched Pop\,II models ($Z = 10^{-6}$, 0.0001, 0.0005, and 0.001). We explored different IMFs for the primary stars (the masses of secondary stars being drawn in a second step; see Sect.~\ref{sec:popIII_reion_GLS}): a classical Chabrier IMF, and three top-heavy IMFs, defined by 
\begin{equation}
    \label{eq;IMF_wise}
    \phi(m) \propto m^{-2.3} \exp\left[- \left( \frac{\mchar}{m} \right) ^{1.6}\right]\,,
\end{equation}
where $\phi(m)dm$ is the number of stars created with masses between $m$ and $m+dm$, and with \mchar = 50\,\Msun, 100\,\Msun and 200\,\Msun. 
We are not interested in the star-formation history over more than a massive-star lifetime and therefore only considered simple stellar population (SSP) models, that is, single coeval stellar populations. Also, we note that adopting stellar populations with binary fractions more typical of nearby star-forming regions \citep[around 70\%,][]{sana2012}, or even as low as in the models by \citet[][around 25\%]{riaz2018}, would negligibly impact our results. This is apparent from the small difference between pure-binary and pure-single star models at the ages up to the emergence of the first SNe (Fig.~\ref{fig:comp_ndot_VLA} below). Even at later ages, which are of less interest to us here, the differences between models including 25\% and 100\% binaries remain modest \citep[see][]{lecroq2024}.

As the first aim of this work is to compare \protect\GLS predictions with the diagnostic diagrams and Pop\,III (single-stars) SEDs presented by \citet{nakajima2022}, we adopted nebular parameters compatible with theirs, that is,
\begin{itemize}
    \item $\zism = Z \in$ \{$10^{-11}$, $10^{-6}$, 0.0001, 0.0005, 0.001\};
    \item $\nh = 10^3 \, \mathrm{cm}^{-3}$;
    \item $\logU = -2$, given that the values considered by \citet{nakajima2022} are between $-3.5$ to $-0.5$;
    \item no dust grains for the two metal-free models.
\end{itemize}
For the Pop\,II models with  $10^{-6} \leq Z \leq 0.001$, we adopted the same values of \CO~=~0.17 and \xid~=~0.3 as in the "standard" models of \citet{lecroq2024} and the same abundance scaling with $Z$ as presented in \citet{gutkin2016}.

\citet{nakajima2022} adopted a primordial He/H abundance of 0.0805, following \citet{hsyu2020}. They also considered a completely metal-free ISM.\footnote{\cloudy offers a "no metals" option, as well as a "no dust grains" option.} We have tested these assumptions against Pop\,III models with metal abundances scaled for $Z = 10^{-11}$ following the method presented by \citet{gutkin2016} and found that the two approaches give very similar results. We therefore chose to adopt the value of \citet{hsyu2020} for the primordial He abundance, and to turn off metals as well as dust grains in \cloudy, for the sake of consistency in our comparison with the work of \citet{nakajima2022}. This choice should have very little impact on the predictions presented in the next sections.

We also tested the impact of adopting a plane-parallel geometry, as assumed by \citet{nakajima2022}, rather than a closed spherical geometry for \cloudy calculations. The results obtained under the two assumptions being extremely similar, we chose not to investigate this aspect further.

We note that, with the assumptions presented here, the only emission-line features relevant to the metal-free models are those related to H and He. We also point out that, as our models do not follow the increase in gas-phase metallicity with stellar population evolution, strictly metal-free models should be considered only up to the time of the explosion of the first SN. This limitation will be discussed further in Sect.~\ref{sec:popIII_res}.

\section{Emission properties of EoR galaxies}
\label{sec:popIII_res}

As Pop\,III stars are expected to be able to trigger intense \lheii-line emission \citep[e.g.,][]{inoue2011, venditti2024b}, we first focus on \lheii-related spectral features and ionizing photons. After discussing \protect\GLS predictions for UV-optical line emission and SEDs, we then look into the predicted H-ionizing photon production rate and its dependence on time and metallicity. Finally, we complete the analysis of the predicted stellar populations by examining other important reionization tracers, such as the LW-band emission, which, as previously discussed, conditions the possibility of star formation in a circumstellar gas cloud, and the statistics of SNe, related to the chemical enrichment in pristine star-forming regions.

\subsection{Ultraviolet and optical emission}
\label{sec:popIII_res_diags}

To investigate the UV and optical emission-line properties of EoR galaxies, we begin by examining the ionizing SEDs predicted by \protect\GLS for populations of single and binary Pop\,III stars. Since stellar effective temperature generally increases with decreasing metallicity (because the reduced opacity makes stars hotter), we expect Pop\,III models to have higher rates of very high-energy photons than more metal-rich ones. For massive stars at extremely low metallicities ($Z \lesssim 10^{-10}$), this phenomenon is amplified by the initial absence of the trace amounts of C, N and O necessary to burn H through the CNO cycle in massive stars. For these extremely low metallicity stars, the only channel for hydrogen fusion is the proton-proton (p-p) process, which is less efficient as a thermostat than the CNO cycle. Thus, gravitational contraction in the late pre-main sequence phase continues until the core reaches the high temperature and density needed to initiate the triple-alpha process (i.e., He burning), which provides the trace amounts of carbon to activate the CNO cycle. The whole structure of massive stars is therefore more compact and hotter at low metallicities \citep{marigo2001,costa2023}.\footnote{In contrast, since low-mass stars are well-supported by the p-p cycle alone, their structure is affected only through the impact of metallicity on opacity.}
Moreover, as the emission is dominated by the most massive stars at early ages, single and binary models should coincide up to ages around 1\,Myr, when the most massive stars leave the main sequence.

\begin{figure}
\centering
\includegraphics[trim=0 0 20 0, clip, width=\columnwidth]{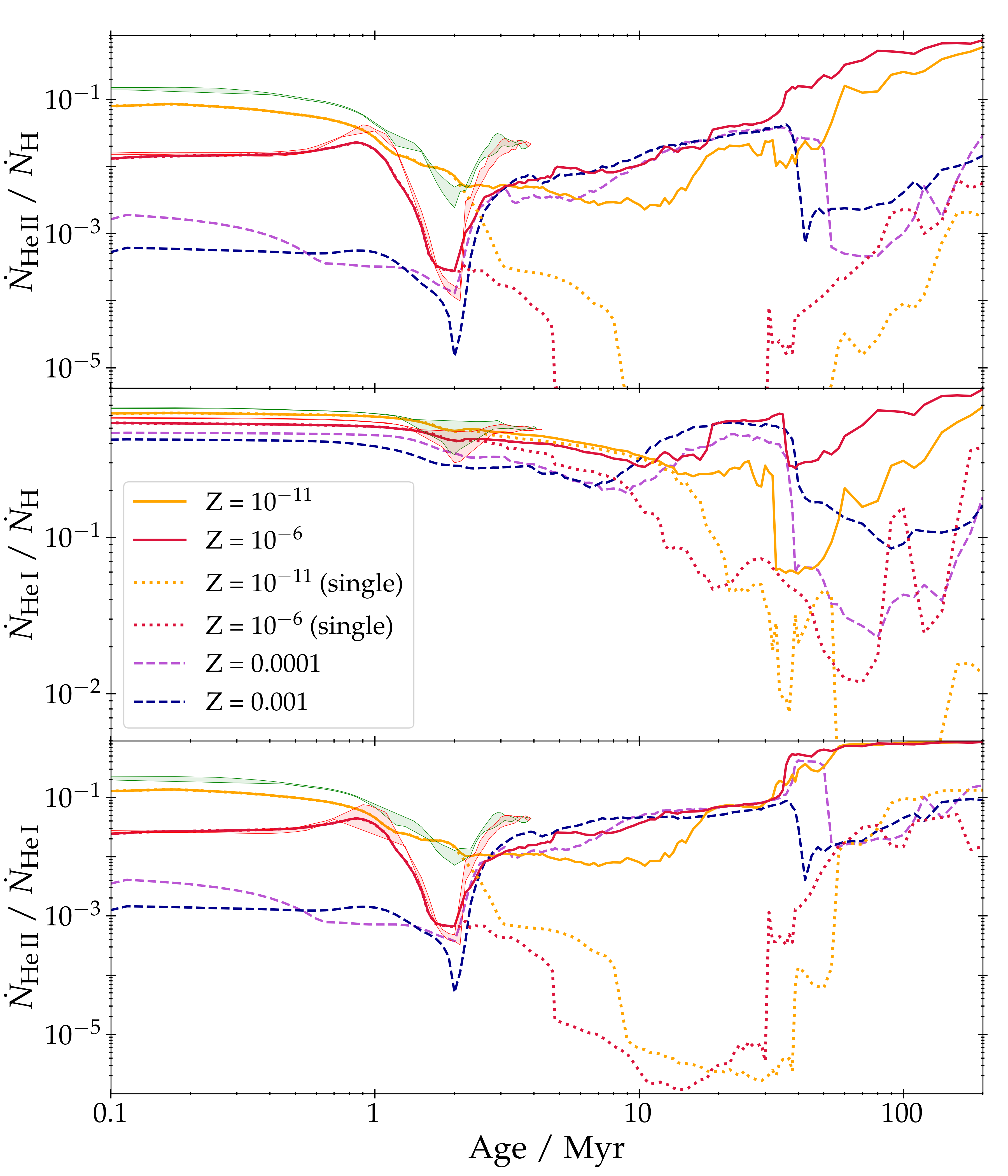}
 \caption{Production rates of \lheii- and \lhei-ionizing photons (normalized to that of H-ionizing photons in the upper two panels and relative to each other in the bottom panel) as a function of the stellar-population age for different \protect\GLS models. Solid orange and red lines denote \protect\GLS Pop\,III ($Z = 10^{-11}$) and extremely metal-poor Pop\,II ($Z = 10^{-6}$), pure-binary models, while dotted lines denote the corresponding single-star models for comparison. Pop\,II pure-binary models for $Z=0.0001$ and 0.001 are shown as dashed purple and blue lines for reference. All these models have the same zero-age \citet{chabrier2003} IMF. The green- and red-shaded regions indicate the areas covered by models with the three top-heavy IMFs with characteristic masses of 50\,\Msun, 100\,\Msun and 200\,\Msun, as described in Sect.~\ref{sec:popIII_reion_params}, for $Z = 10^{-11}$ and $10^{-6}$, respectively. These models are considered only up to the point where their absolute UV magnitude reaches $M_{\mathrm{UV}} =- 14$, beyond which the stars are too few and the curves too noisy \citep[see][for more details]{lecroq2024}.}
\label{fig:comp_ndot_VLA}
\end{figure}

Figure~\ref{fig:comp_ndot_VLA} shows the production rates of \lheii- and \lhei-ionizing photons, normalized to that of H-ionizing photons, as a function of the stellar-population age for different \protect\GLS models. We show models of Pop\,III stellar populations ($Z = 10^{-11}$, represented in orange) and extremely metal-poor Pop\,II stellar populations ($Z=10^{-6}$, in red), with solid lines for pure-binary and dashed lines for single-star models. Pop\,II models with $Z = 0.0001$ and 0.001 are represented as dotted lines (the model with $Z = 0.0005$, not shown for clarity, would lie between these two models). These models all have the same zero-age \citet{chabrier2003} IMF. As expected from the above argument, models with $Z = 10^{-11}$ and $10^{-6}$ have much higher $\dot{N}_{\mathrm{He\,II}}$ than models with metallicities of a few percent of solar at early ages, due to their main sequence (MS) being shifted to higher temperatures.\footnote{While the MS of $Z\ga0.0001$ models roughly spans from 20,000 to 60,000\,K, it is shifted to the 40,000--80,000\,K range for $Z = 10^{-6}$ and 45,000--100,000\,K for $Z = 10^{-11}$.} Figure~\ref{fig:comp_ndot_VLA} also confirms that the difference between single- and binary-star models appears after the first million years of evolution (as already highlighted in figure~1 of \citealt{lecroq2024}). Then, around 2\,Myr, both Pop\,III and Pop\,II binary-star models exhibit a high $\dot{N}_{\mathrm{He\,II}}$, which is absent in single-star models. This arises from the formation of pure-He, WNE-like products of massive-star stripping. However, it is interesting to note that this increase in $\dot{N}_{\mathrm{He\,II}}$ is modest for $Z = 10^{-11}$ and becomes more pronounced as metallicity increases. This is due to the more rapid evolution of extremely low metallicity stars, because of their accelerated and hotter central-H burning: the WNE-like products of the evolution of the most massive stars begin to produce ionizing photons before most of the lower-mass stars leave the MS.
Finally, another notable difference arises at ages greater than $\sim40$\,Myr, where the production-rate ratios of high-energy photons drops in Pop\,II models before rising again. This is due to the contribution from accretion disks of XRBs, which grows at earlier ages in Pop\,III models due to the faster stellar evolution.\footnote{Pop\,III models not including the emission from XRB accretion disks do not exhibit this feature. Instead, they show energetic-photon production rates which fall off more rapidly than in Pop\,II models, due to their faster evolution and their lack of WNE-type stars, present at late ages in Pop\,II models.} This feature, consistent with the growing impact of XRB accretion disks with population age as suggested in previous studies (see Section~4.3 of \citealt{lecroq2024}), occurs at ages beyond the primary focus of the present study.

These differences in behavior between Pop\,III and Pop\,II models, as well as between single and binary stars, are less marked and appear later in the case of $\dot{N}_{\mathrm{He\,I}}$. This is because the conditions to produce photons capable of singly ionizing He are less extreme -- the first ionization energy of helium being 24.6\,eV, less than half the 54.4\,eV required to doubly ionize it.

The green- and red-shaded regions show the areas covered by models with the three top-heavy IMFs with characteristic masses of 50\,\Msun, 100\,\Msun and 200\,\Msun described in Sect.~\ref{sec:popIII_reion_params}, also with metallicities of $Z = 10^{-11}$ and $10^{-6}$ respectively. Their evolution is shown until their absolute UV magnitude reaches $M_{\mathrm{UV}} =- 14$, at which point only few stars remain, and the models become stochastically dependent on the seed used to draw their initial properties, as described by \citet[][see their figure~10]{lecroq2024}. The $Z = 10^{-11}$ model exhibits a globally higher $\dot{N}_{\mathrm{He\,II}}$ with this choice in IMF, resulting from a large fraction of extremely massive stars. This tendency is somewhat less strong for the slightly cooler $Z = 10^{-6}$. Once again, the evolution of $\dot{N}_{\mathrm{He\,I}}$ is very similar to that of models with a \citet{chabrier2003} IMF, since the difference in the fraction of very massive stars does not influence much the production rate of medium-energy ionizing photons.

We now consider the zero-metallicity stellar-population models of \citet{schaerer2003}, based on an earlier version of the Padova evolutionary tracks for single zero-metallicity stars with negligible mass loss, complemented with other models for stars with high mass loss \citep{lejeune2001,meynet2002}, both sets covering the mass range from 1\,\Msun to 500\,\Msun. These stellar tracks do not include binary interactions or the effects of rotation. These models are provided for a range of modified \citet{salpeter1955} IMFs with different lower- and upper-mass cut-offs in the interval between 1 and 500\,\Msun, as described by \citet{raiter2010}. We compare the predictions of these models with those of standard \protect\GLS models with a \citet{chabrier2003} IMF truncated at 0.1 and 300\,\Msun for primary stars and a \citet[][]{sana2012} distribution of secondary-to-primary mass ratios at age zero (single stars being modeled as noninteracting binaries). Such an IMF can lead in binary-star populations to the presence of merged stars with masses up to $\sim600\,\Msun$. For completeness, we investigated the impact of adopting an upper mass cut-off of 100\,\Msun instead of 300\,\Msun. We found the effect on $\dot{N}_{\mathrm{He\,II}}/\dot{N}_{\mathrm{H}}$ and $\dot{N}_{\mathrm{He\,I}}/\dot{N}_{\mathrm{H}}$ to be limited, due to the very small fraction of stars in the upper part of \citet{chabrier2003} distribution, and the extremely short lifetime of these very massive stars. Moreover, low-metallicity stars with masses up to $\sim300\,\Msun$ have been observed in local environments \citep[e.g.,][see also \citealt{vink2011}]{crowther2016,smith2016}, and are expected to become more frequent at lower metallicities and higher redshift. Since in any case Pop\,III stars are expected to be very massive (up to $\sim1000\,\Msun$ for Pop\,III stars), we chose not to investigate this aspect further here and fix the upper mass cut-off of \protect\GLS models at 300\,\Msun for the remainder of this discussion.

\begin{figure}
\centering
\includegraphics[trim=0 0 20 0, clip, width=\columnwidth]{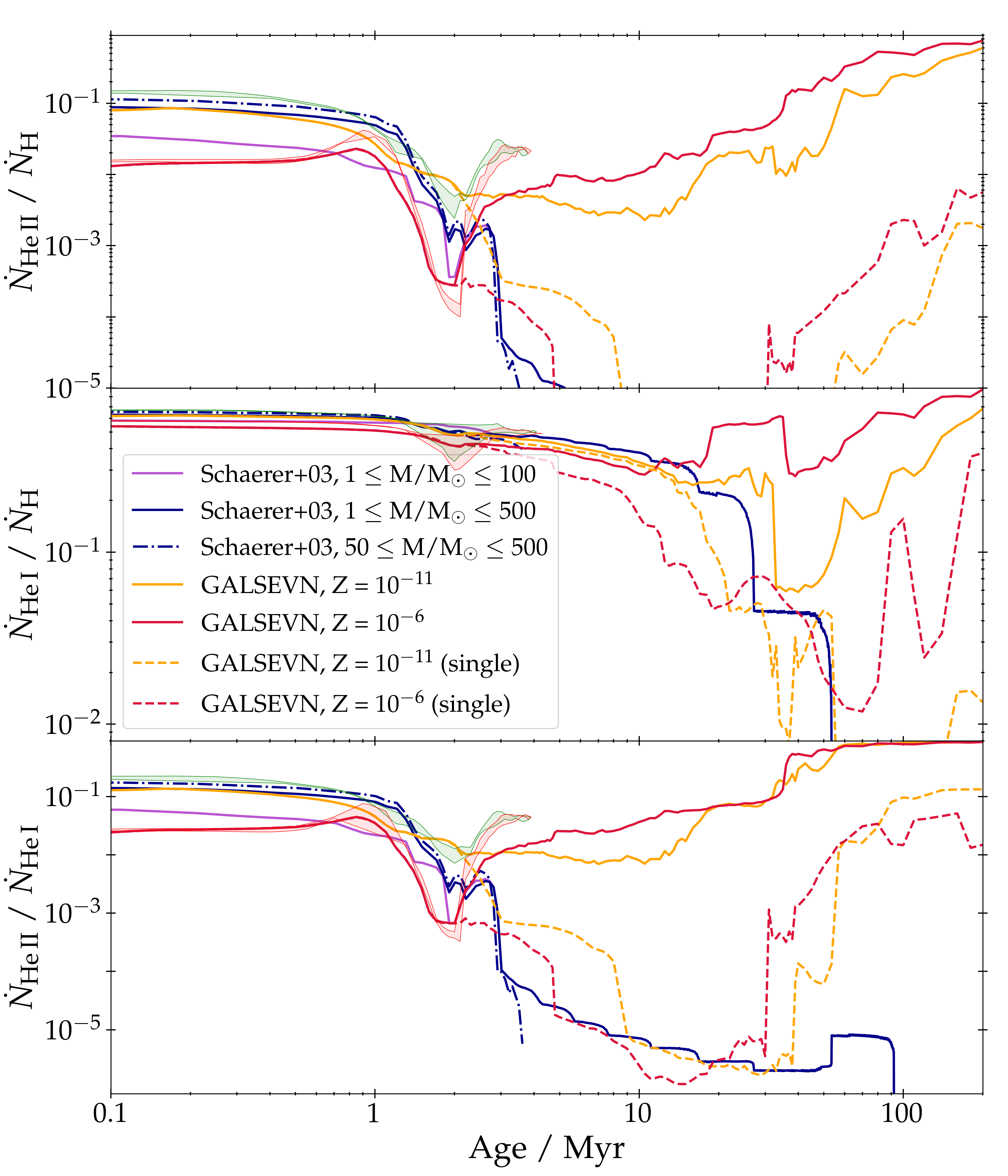}
 \caption{Same as Fig.~\ref{fig:comp_ndot_VLA}, but comparing the prediction of \protect\GLS and \citet{schaerer2003} Pop\,III models. The solid- and dashed-orange and red curves represent the same \protect\GLS $Z=10^{-11}$ and $10^{-6}$ models as in Fig.~\ref{fig:comp_ndot_VLA}, with a \citet{chabrier2003} IMF truncated at 0.1 and 300\,\Msun for primary stars, combined with a \citet[][]{sana2012} distribution of secondary-to-primary mass ratios at age zero (with single stars modeled as noninteracting binaries). The green- and red-shaded regions indicate the areas covered by models with the same three top-heavy IMFs with characteristic masses of 50\,\Msun, 100\,\Msun and 200\,\Msun as  in Fig.~\ref{fig:comp_ndot_VLA}. The solid purple, solid blue and dot-dashed blue curves show the zero-metallicity, single-star models of \citet{schaerer2003} for three \citet{salpeter1955} IMFs with different lower- and upper-mass cut-offs.}
\label{fig:comp_ndot_schaerer}
\end{figure}

Figure~\ref{fig:comp_ndot_schaerer} compares the ionizing SEDs of \citet{schaerer2003}  Pop\,III models for different IMFs with the \protect\GLS $Z=10^{-11}$ and $10^{-6}$ predictions presented in Fig.~\ref{fig:comp_ndot_VLA}. The evolution of $\dot{N}_{\mathrm{He\,II}}$ and $\dot{N}_{\mathrm{He\,I}}$ is quite similar between the standard model of \citet{schaerer2003}, which uses a \citet{salpeter1955} IMF truncated at 1 and 500\,\Msun (in dark purple), and the \protect\GLS single-star Pop\,III ($Z = 10^{-11}$) model (dashed orange). Slight differences in $\dot{N}_{\mathrm{He\,II}}$ between 1 and 2\,Myr likely stem from differences in the stellar tracks and spectral libraries used in the two models.

The model by \citet{schaerer2003} with a \citet{salpeter1955} IMF truncated at 1 and 100\,\Msun exhibits a $\dot{N}_{\mathrm{He\,II}}$/$\dot{N}_{\mathrm{H}}$ ratio about five times lower than those with an IMF reaching 500\,\Msun at the earliest ages. This difference becomes less pronounced as the most massive stars leave the MS phase. 
In the \citet{schaerer2003} model with a \citet{salpeter1955} IMF truncated at 50 and 500\,\Msun, the $\dot{N}_{\mathrm{He\,II}}/\dot{N}_{\mathrm{H}}$ and $\dot{N}_{\mathrm{He\,I}}/\dot{N}_{\mathrm{H}}$ ratios are initially slightly boosted relative to the other two models by the removal of the of the coolest ionizing stars, although they remain below the predictions of the \protect\GLS models with top-heavy IMFs.

Based on these preliminary considerations regarding the \citet{schaerer2003} SEDs used by \citet{nakajima2022} to predict the properties of EoR galaxies, it is now interesting to compare \protect\GLS predictions with the diagnostics these authors propose to differentiate ionization by Pop\,III stars from that by other high-energy sources in emission-line diagrams: pristine, direct-collapse black holes (DCBHs),\footnote{DCHBs are black holes with masses between $10^5$ and $10^6$ \Msun, which might have formed directly from the collapse of pristine gas clouds in the early Universe \citep[see, e.g.,][for reference]{volonteri2012,valiante2016,beckmann2019,inayoshi2020}.} evolved Pop\,II stars and metal-enriched AGNs. \citet{nakajima2022} considered theoretical DCBH SEDs composed of a black-body-like spectrum peaking in the ultraviolet combined a power-law tail in the X-ray range, similar to AGN emission \citep{valiante2018}, with adjustable black-body temperature and spectral index of the power law. They also considered Pop\,III stars and DCBHs evolving in an enriched medium of adjustable metallicity. \citet{nakajima2022} computed the ionizing spectra of Pop\,II galaxies using \bpass binary-star models with metallicities ranging from $10^{-5}$ to $10^{-3}$, for 10\,Myr-old stellar populations with constant SFR. 
They modeled the ionizing radiation from AGNs using the same templates as for DCBHs but with higher metallicities. They then used the \cloudy photoionization code to compute the nebular emission produced by these different sources, with adjustable parameters similar to those detailed in Sect.~\ref{sec:popIII_reion_params}.

In their paper, \citet{nakajima2022} investigated several emission-line diagrams able to isolate the signatures of either Pop\,III stellar populations or pristine DCBHs. They conclude that ionization by Pop\,III stars can be best discriminated based on the equivalent widths of \heiiopt and, to a lesser extent, \heii. Extreme values of these equivalent widths cannot be reproduced by any other source in their models, including primeval DCBHs. They also find the line ratios \heiiopt/\hb and \heii/\lya useful to identify populations of Pop\,III stars with a top-heavy IMF, although these ratios do not discriminate between zero-metallicity stars and DCBHs. \citet{nakajima2022} also show that diagrams involving \lhei lines are not good indicators of metallicity, since the values for Pop\,II and Pop\,III stellar populations are similar for these ratios. They finally identify three criteria and tendencies based on these emission lines to identify Pop\,III stars in emission-line diagrams (see their figures~2, 3, and 6).

\begin{figure}
\centering
\includegraphics[width=0.83\columnwidth]{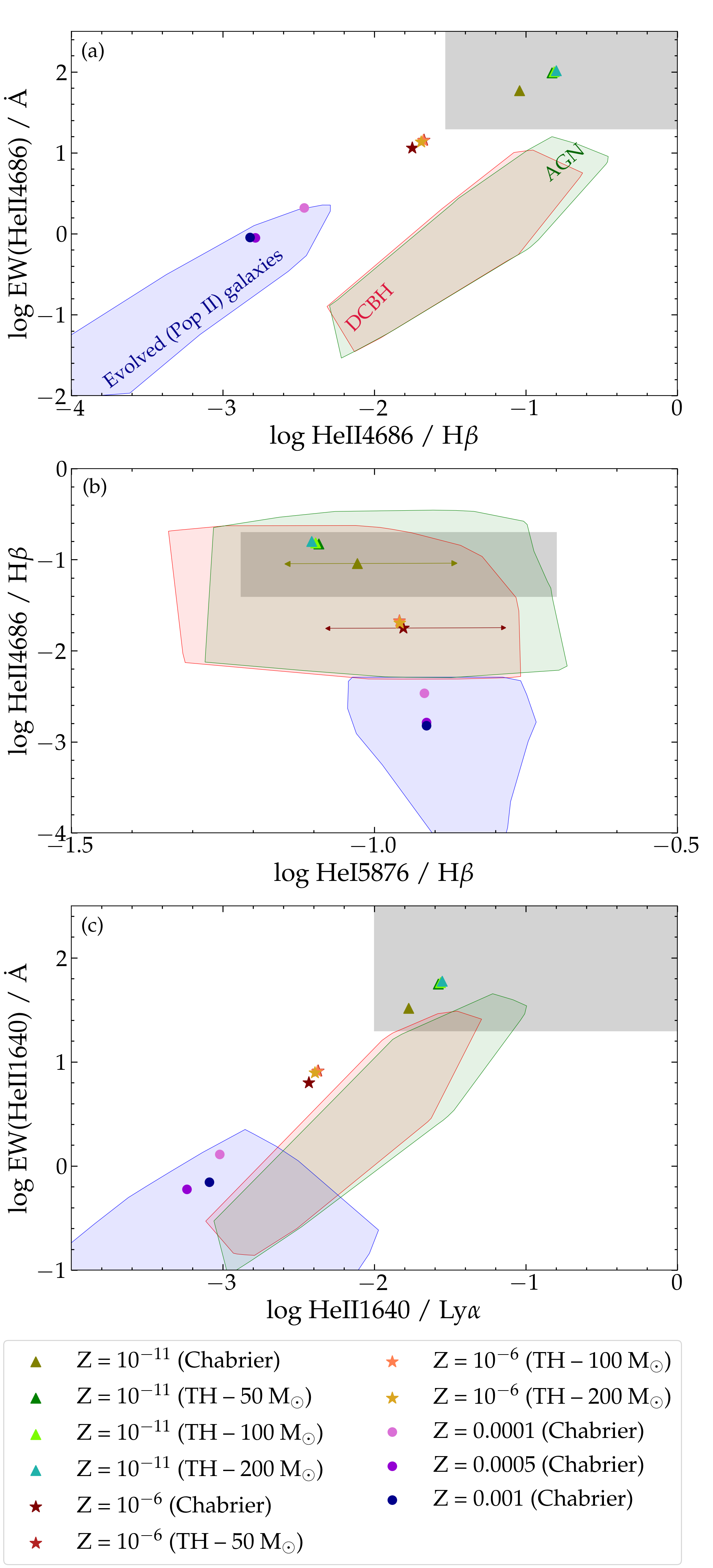}
 \caption{Diagnostic diagrams proposed by \citet{nakajima2022} to differentiate ionization by Pop\,III stars from that by other high-energy sources. The markers represent zero-age, pure binary-star \protect\GLS models with different metallicities and IMFs, color-coded as indicated on the bottom. The gray-shaded area denotes the region identified by \citet{nakajima2022} as populated by Pop\,III stars, while the other colored regions correspond to areas populated by primeval DCBHs, evolved (Pop\,II) galaxies, and AGNs, as indicated. In panel~(b), the horizontal arrows for $Z = 10^{-11}$ and $10^{-6}$ indicate the positions of corresponding models with \nh = $10^{2}$ cm$^{-3}$ (left-pointing) and $10^{4}$ cm$^{-3}$ (right-pointing).}
\label{fig:comp_1st-age_zones}
\end{figure}

\begin{figure}
\centering
\includegraphics[trim=0 0 0 40, clip, width=1.04\columnwidth]{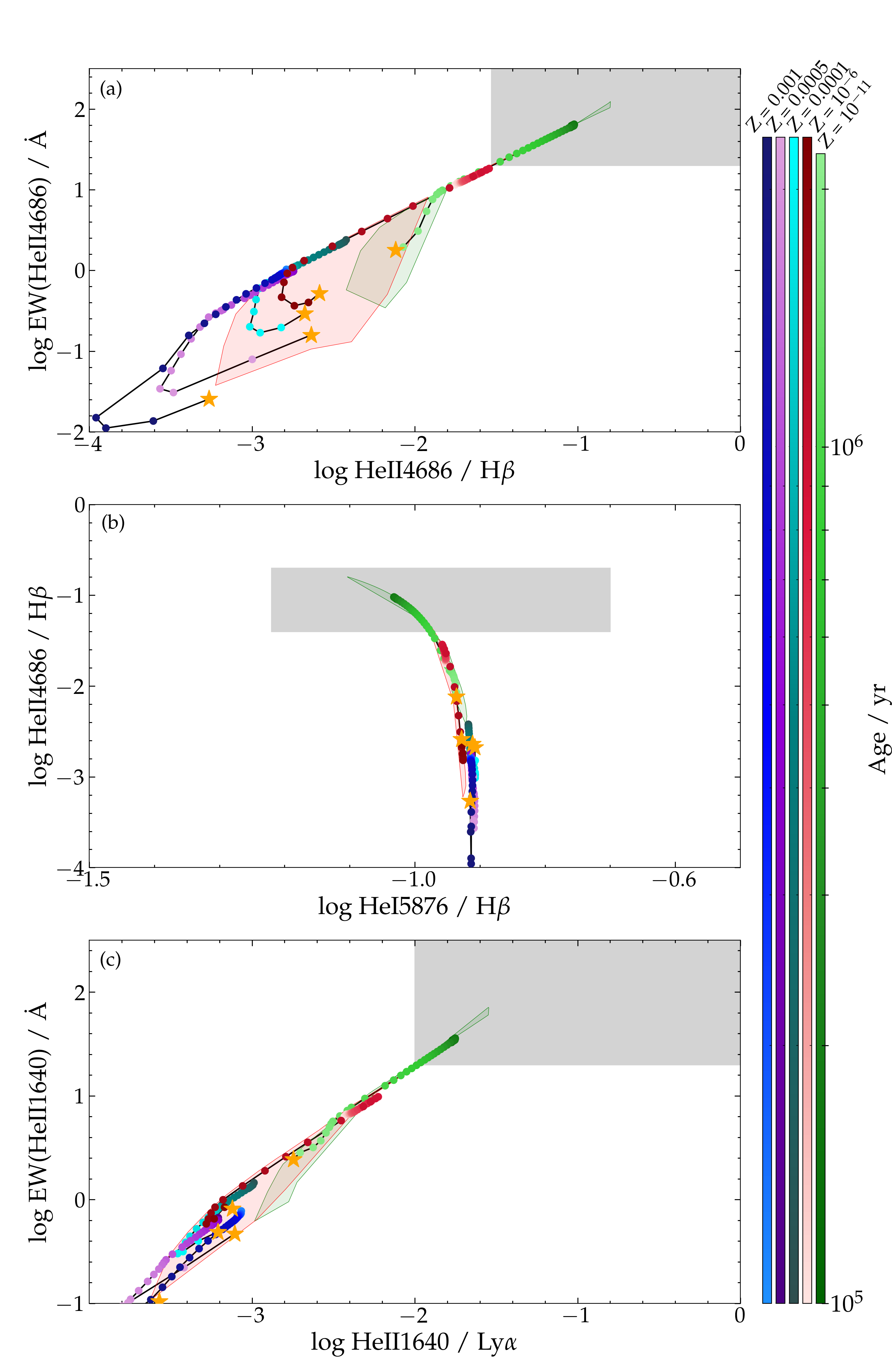}
 \caption{Same diagnostic diagrams as in Fig.~\ref{fig:comp_1st-age_zones}. As before, the gray-shaded zone in each diagram indicates the location of models powered by Pop\,III stars, as identified by \citet{nakajima2022}. The different curves with dots show the temporal evolution of the \protect\GLS SSP models presented at zero age in Fig.~\ref{fig:comp_1st-age_zones}, with age referenced by the color bars on the right. For each curve, an orange star marks the appearance of the first SN, at which point the model is stopped in the present study. The red- and green-shaded regions show the areas covered by models with top-heavy IMFs, as in Figs.~\ref{fig:comp_ndot_VLA} and \ref{fig:comp_ndot_schaerer}. }
\label{fig:comp_to_SN}
\end{figure}

We plot these emission-line diagrams and the location of the \citet{nakajima2022} models in Fig.~\ref{fig:comp_1st-age_zones}, with the region corresponding to Pop\,III stars shown in shaded gray. Although the diagram represented in Fig.~\ref{fig:comp_1st-age_zones}b) is not strictly identified as a diagnostic diagram, it is also useful as it can be used to discriminate between several values of \nh.
The model predictions presented by \citet{nakajima2022} for DCBHs, evolved Pop\,II galaxies, and AGNs -- over their whole range of gas-phase metallicity, \logU, \nh, and IMFs -- appear in this figure as color-shaded regions (respectively in red, blue, and green). 
The colored dots in Fig.~\ref{fig:comp_1st-age_zones} show the locations of zero-age\footnote{In the \citet{nakajima2022} models, all stars begin their evolution on the zero-age main sequence. Instead, in the \GLS model, which includes pre-MS evolution, SSPs ignite when the most massive stars approach the MS. This difference in timing is minimal, as pre-MS evolution lasts only about 1\% of the MS duration, and does not impact the conclusions of this paper.} \protect\GLS SSP models with different metallicities and IMFs. 

Figure~\ref{fig:comp_1st-age_zones} shows that the \protect\GLS predictions for $Z = 10^{-11}$ strongly support the criteria proposed by \citet{nakajima2022} to identify signatures of Pop\,III stellar populations. These criteria appear stringent enough that even Pop\,II models with $Z = 10^{-6}$ fall below the identified thresholds. This agreement is not surprising, as Fig.~\ref{fig:comp_ndot_schaerer} shows a similarity in the ionizing SEDs of the models, especially at early ages, where binaries are not a decisive factor. The most effective diagnostic diagram appears to be EW(\heiiopt) versus \heiiopt/\hb, as it clearly separates regions populated by evolved galaxies, AGNs, DCBHs, and zero-metallicity models. The equivalent ultraviolet diagram of EW(\heii) versus \heii/\lya also distinguishes these sources effectively, although there is some overlap in the lowest-EW(\heii) part of the gray-shaded region. In contrast, the pure line-ratio diagram of \heiiopt/\hb versus \hbox{He\,\textsc{i}$\lambda5876$}/\hb separates evolved galaxies from all other sources effectively, but struggles to differentiate AGNs and DCHBs from zero-metallicity stars. This diagram does, however, show a strong dependence on \nh, as indicated by the horizontal arrows pointing to the locations of equivalent models with $\nh = 10^{2}$ and $10^{4}$\,cm$^{-3}$, for the lowest two metallicities. Thus, while this diagram does not fully resolve all sources, it provides useful complementary information for Pop\,III diagnostics.

We note that the predictions of the \protect\GLS model for evolved stellar populations with $Z = 0.0001$, 0.0005, and 0.001 are also in good agreement with those of \citet{nakajima2022}. This is not surprising, given the agreement between \protect\GLS and \bpass predictions at early ages \citep[see figure~4 of][]{lecroq2024}.

We did not investigate predictions for Pop\,III stars evolving in a slightly enriched ISM, as do some of the models presented by \citet{nakajima2022}. Although our approach would allow such modeling \citep[see section~3.3.3 of][]{lecroq2024}, we made this choice because of the short lifetimes of these stars, and because we focus in this study on the very first generations of Pop\,III stars.

It is of interest to examine the predicted signatures of primeval sources not only at zero age, but also in their evolution with time, to be able to estimate the probability of observing them in the areas of diagnostic diagrams identified above. We therefore also discuss the time evolution of these spectral features for the \protect\GLS models commented above. Again, the \protect\GLS models presented here do not take into account the enrichment of the surrounding medium through stellar evolution, but assume a constant gas-phase metallicity equal to that of the ionizing stars. Such enrichment could considerably alter the values of \zism for the lowest metallicities, which would be highly sensitive to the presence of even trace amounts of metals. We therefore stopped our calculations at the time when the first SN appeared for each population, to ensure that the surrounding medium would not have yet been polluted by new metals.

Figure~\ref{fig:comp_to_SN} shows the time evolution of the same \protect\GLS SSP models as displayed at zero age in Fig.~\ref{fig:comp_1st-age_zones}. For clarity, the evolution has been detailed only for models with a \citet{chabrier2003} IMF, the areas sampled by models with top-heavy IMFs appearing as green- and red-shaded regions for the metallicities $Z = 10^{-11}$ and $10^{-6}$, respectively (while regions identified by \citealt{nakajima2022} as populated by primeval DCBHs, Pop\,II galaxies, and AGNs have been omitted). In each model, an orange star marks the time of appearance of the first SN, when the model is stopped, corresponding to about 2.3\,Myr for all models. 
Fig.~\ref{fig:comp_to_SN} reveals that the evolution of all considered spectral features is very rapid at these early ages. Indeed, models with $Z = 10^{-11}$ remain within the gray zones identified above to select ionization by Pop\,III stars only for ages up to $\sim1$\,Myr. This makes the probability of observing Pop\,III stars in this state quite low. 
Models with higher metallicities exhibit loops toward stronger \heii spectral features, around 1\,Myr and later after 2\,Myr. These loops, especially visible for the \heii equivalent widths predicted for $Z = 10^{-6}$, are due to the sudden increase in the production rate of \lheii-ionizing photons already discussed in Fig.~\ref{fig:comp_ndot_VLA}. However, the implied increases in EW(\heii), EW(\heiiopt), and \heiiopt/\hb are not intense enough to bring these models over the thresholds of the Pop\,III diagnostic criteria.

Figures~\ref{fig:comp_1st-age_zones} and \ref{fig:comp_to_SN} overall highlight the fact that the \protect\GLS models support the criteria presented by \citet{nakajima2022} to identify ionization by Pop\,III stellar populations. The figures also confirm that these criteria as strict enough to efficiently separate Pop\,III signatures from those from other possible sources of high-energy photons. However, the high thresholds of these criteria make them accurate for only a very short time, making the probability of actually observing stellar populations in this regime very low.

\subsection{Production efficiency of ionizing photons}
\label{sec:popIII_res_xiion}

A crucial parameter to constrain the amount of H-ionizing photons available for reionization in simulations is the Lyman-continuum escape fraction, \fesc, defined as the fraction of all ionizing photons produced that can escape through the ISM and IGM to ionize intergalactic neutral hydrogen. Recently, significant efforts have been made to directly measure \fesc at redshifts up to $z\sim4$. However, because of the substantial increase in IGM opacity at higher redshift \citep[e.g.,][]{inoue2014}, \fesc in the EoR can be inferred only indirectly and through model-dependent approaches. To address this limitation, indicators need to be established based on local observations, which can link readily measurable quantities, such as line emission (which traces the fraction of ionizing photons absorbed by the gas) and nonionizing UV luminosity. When combined with the production efficiency of ionizing photons, \xiion, which links the production rate of H-ionizing photons to that of nonionizing ultraviolet photons, such observations may provide more confident estimates of \fesc at high redshift. 

 Three different definitions of the production efficiency of ionizing photons can be found in the literature, each with varying degrees of appropriateness depending on whether the focus is on a theoretical or observational study:
\begin{itemize}
    \item the stellar ionizing-photon production efficiency,
    \begin{equation}
        \label{eq:xiionst}
        \xiionst = \dot{N}_{\mathrm{ion}} / L^{*}_{\mathrm{UV}}\,,
    \end{equation}
    where $L^{*}_{\mathrm{UV}}$ is the intrinsic stellar monochromatic ultraviolet luminosity, that is, the UV luminosity that would be observed in the absence of gas and dust in the galaxy; 
    \item the stellar+nebular ionizing-photon production efficiency, 
    \begin{equation}
        \label{eq:xiionneb}
        \xiionHII = \dot{N}_{\mathrm{ion}} / L^{\mathrm{HII}}_{\mathrm{UV}} \,,
    \end{equation}
    where $L^{\mathrm{HII}}$ is the UV luminosity accounting for the effects of dust absorption within \hii regions and nebular recombination-continuum emission; 
    \item and the observed ionizing-photon production efficiency, 
    \begin{equation}
        \label{eq:xiionobs}
        \xiion = \dot{N}_{\mathrm{ion}} / L_{\mathrm{UV}}\,,
    \end{equation}
    where $L_{\mathrm{UV}}$ is the observed, uncorrected UV luminosity.
\end{itemize}
These monochromatic UV luminosities are computed from the stellar spectrum, the stellar+nebular transmitted spectrum, and the rest-frame observed spectrum, respectively, by averaging the emission over a 100\,\AA{}-wide window centered on 1500\,\AA\ \citep[see, e.g.,][]{robertson2013}.

To circumvent the difficulty posed by uncertainties over \fesc values as explained above, we focused more particularly on two limiting cases, studying both the stellar coefficient \xiionst -- which corresponds to the case where \fesc = 1, meaning that all ionizing photons escape from the surrounding \hii region -- and the nebular coefficient \xiionHII -- corresponding to \fesc = 0, as we are considering an ionization-bounded \hii region.
The former depends only on stellar-population properties, that is, mainly age, metallicity, and IMF. Simulations often consider a fixed value of \xiionst, constant with age and chemical evolution. One of the goals of the present work is to provide an overview of \xiionHII and \xiionst for \protect\GLS SEDs, including how they evolve with these stellar-population parameters, which we can achieve self-consistently with the ultraviolet and optical spectral signatures mentioned above.

The nebular ionizing-photon production efficiency \xiionHII is a more direct observable than \xiionst, as it can be constrained by observations corrected for dust attenuation outside the ionized region. We therefore begin our analysis by looking at the evolution of \xiionHII with time and metallicity, for the SSP models discussed in the previous section. The quantity \xiionHII was calculated using the rate of H-ionizing photons predicted by the \protect\GLS stellar-population models, and the UV luminosity computed from the spectrum output by \cloudy, to account for the recombination continuum and dust absorption inside the ionized region.

\begin{figure}
\centering
\includegraphics[width=\columnwidth]{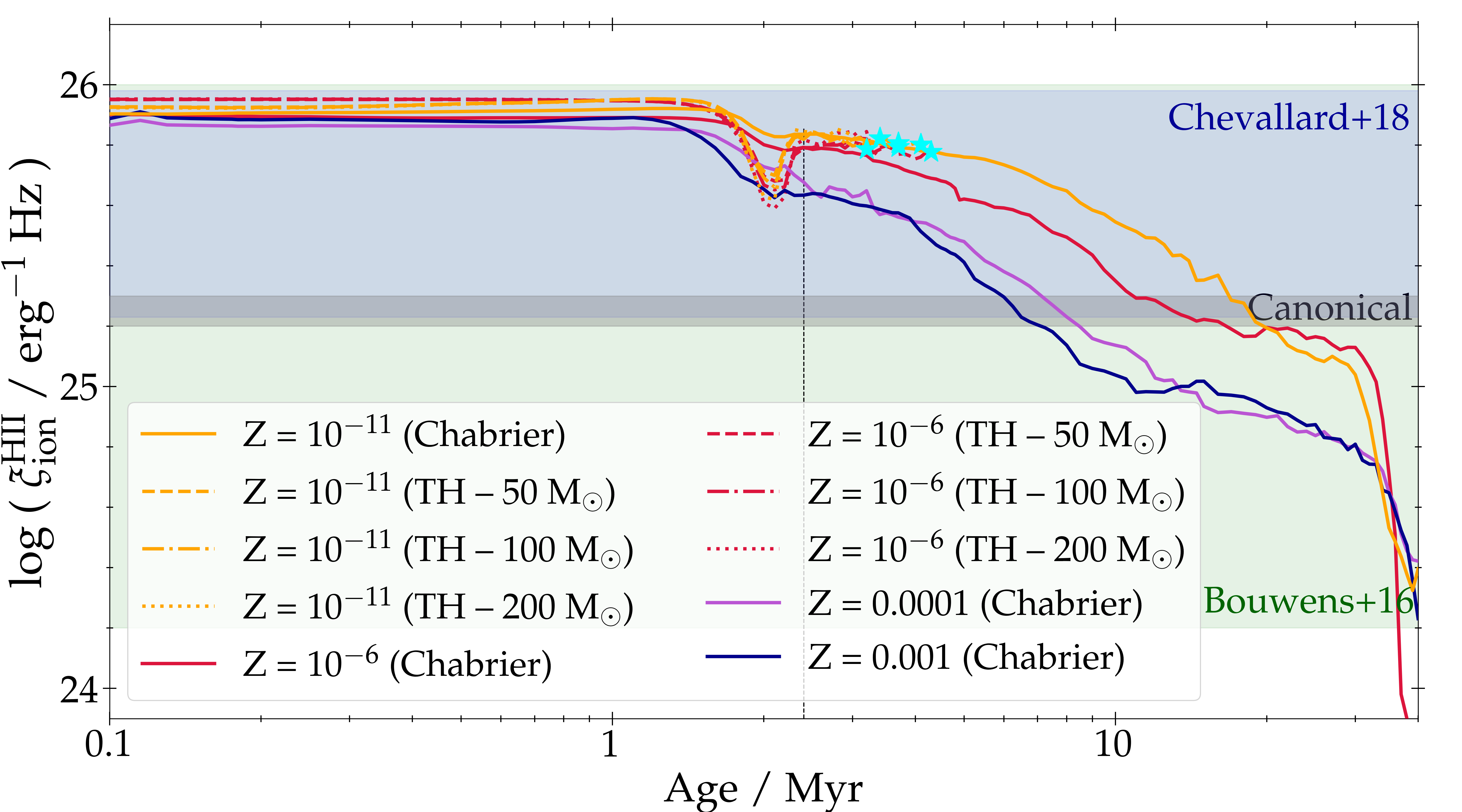}
 \caption{Nebular ionizing-photon production efficiency \xiionHII as a function of the stellar-population age for the same \protect\GLS models as in Fig.~\ref{fig:comp_ndot_VLA}. The time evolution is followed up to 40\,Myr, but the \hii regions surrounding the young stellar populations are expected to have been disrupted before at most 10\,Myr \citep{murray2011,ma2015}. The vertical dashed line marks the time of appearance the first SN in the lowest-metallicity models, corresponding to the beginning of chemical enrichment. The models with top-heavy IMFs are plotted until their absolute UV magnitude reaches $M_{\mathrm{UV}} =- 14$, marked by cyan stars. The blue and green-shaded regions correspond to \xiionHII constraints derived from observations at low to intermediate redshifts, from different works. The gray-shaded region corresponds to the canonical values usually adopted for \xiionst in simulations.}
\label{fig:evo_xiionneb}
\end{figure}

Figure~\ref{fig:evo_xiionneb} shows the time evolution of \xiionHII for the same \protect\GLS binary SSP models as in Fig.~\ref{fig:comp_ndot_VLA}. 
The time of appearance of the first SN in the lowest metallicity populations is marked by the vertical dashed line, to indicate the beginning of chemical enrichment. Estimates of \xiion have flourished in recent work, both from simulations and from observational constraints. Analytical reionization models often assume a canonical value in the range $\log\, (\xiionst / \mathrm{erg}^{-1} \mathrm{Hz}) \approx 25.2$--$25.3$ \citep[e.g.,][]{robertson2013}, shown by the gray-shaded region in Fig.~\ref{fig:evo_xiionneb}. We can also compare \protect\GLS predictions to the \xiion ranges estimated in two recent observational studies. \citet{bouwens2016} derived their values from observations of star-forming galaxies in the GOODS field, at redshifts $z\sim3.8$--$5.4$, based on the measured UV-continuum slope and \ha intensities. \citet{chevallard2018} obtained their values by fitting the SEDs of ten nearby analogs of primeval galaxies from \HST/COS observations, using the \beagle spectral-interpretation tool. The ranges of \xiion values in these two studies appear as green- and blue-shaded regions on Fig.~\ref{fig:evo_xiionneb}. These results are in general agreement with those derived in several other recent works, which used different methods to retrieve the values of \xiionst and \xiionHII from observations at various redshifts ($0.3\,\lesssim z \lesssim\,9$), and whose ranges of values have not been plotted in Fig.~\ref{fig:evo_xiionneb} for the sake of clarity \citep[e.g.,][see section~5 of \citealt{chevallard2018} for a more detailed comparison]{stark2015,schaerer2016,izotov2017b,matthee2017,stark2017,shivaei2018}.

Figure~\ref{fig:evo_xiionneb} shows that the \protect\GLS predictions are very similar for all metallicities at early ages, with a relatively high estimated \xiionHII (still falling within the observed range), which remains almost constant at first, while most stars evolve on the MS. After $\sim1$\,Myr, the dependence of the evolution becomes more pronounced. The lowest-metallicity models generally keep a higher \xiionHII, and the decrease in \xiionHII seems to be almost linear with slopes depending on metallicity until $\sim30$\,Myr. The slope seems to steepen with increasing metallicity. This is expected, due to the higher ionizing power of stars with extremely low metallicity, resulting from their specific properties (e.g., evolutionary tracks shifted toward higher temperatures), as discussed in the previous section.

Moreover, comparison of the models with a \citet{chabrier2003} and top-heavy IMFs reveals that the IMF does not appear to be a crucial parameter for the predicted \xiionHII values. While the predicted values for top-heavy IMFs are slightly higher at early ages, the main difference is a pronounced feature just before 2\,Myr, when most massive stars reach the end of their H-burning phase, resulting in a drop in the production of ionizing photons until the appearance of hot, WNE-like products of binary evolution. As before, we stop the evolution of models with top-heavy IMFs when their absolute UV magnitude reaches $M_{\mathrm{UV}} =- 14$.

Figure~\ref{fig:evo_xiionneb} shows overall good agreement between \protect\GLS predictions for \xiionHII and recent observations. It also reveals a dependence of \xiionHII on time and metallicity, which is almost negligible for approximately the first million years, before becoming more pronounced.

According to simulations of early galaxy formation \citep[e.g.,][]{finkelstein2019} and recent observational constraints on the nonionizing UV luminosity density \citep[e.g.,][]{atek2024}, with such high values of \xiionHII, the escape of only a few percent of ionizing photons from nascent galaxies would suffice to reionize the Universe. The fact that such high values of \xiionHII are naturally produced by \protect\GLS models for metallicities up to a few percent of solar supports the major role that early star-forming galaxies are likely to have played as reionization drivers.

For some applications to the modeling of early galaxy formation, it may be useful to have a simple analytical expression for estimates of \xiionHII as a function of the time and metallicity. We derive such relations here for models with a \citet{chabrier2003} IMF (as Fig.~\ref{fig:evo_xiionneb} shows, models with top-heavy IMFs have similar properties at early ages). 

\begin{figure}
\centering
\includegraphics[width=\columnwidth]{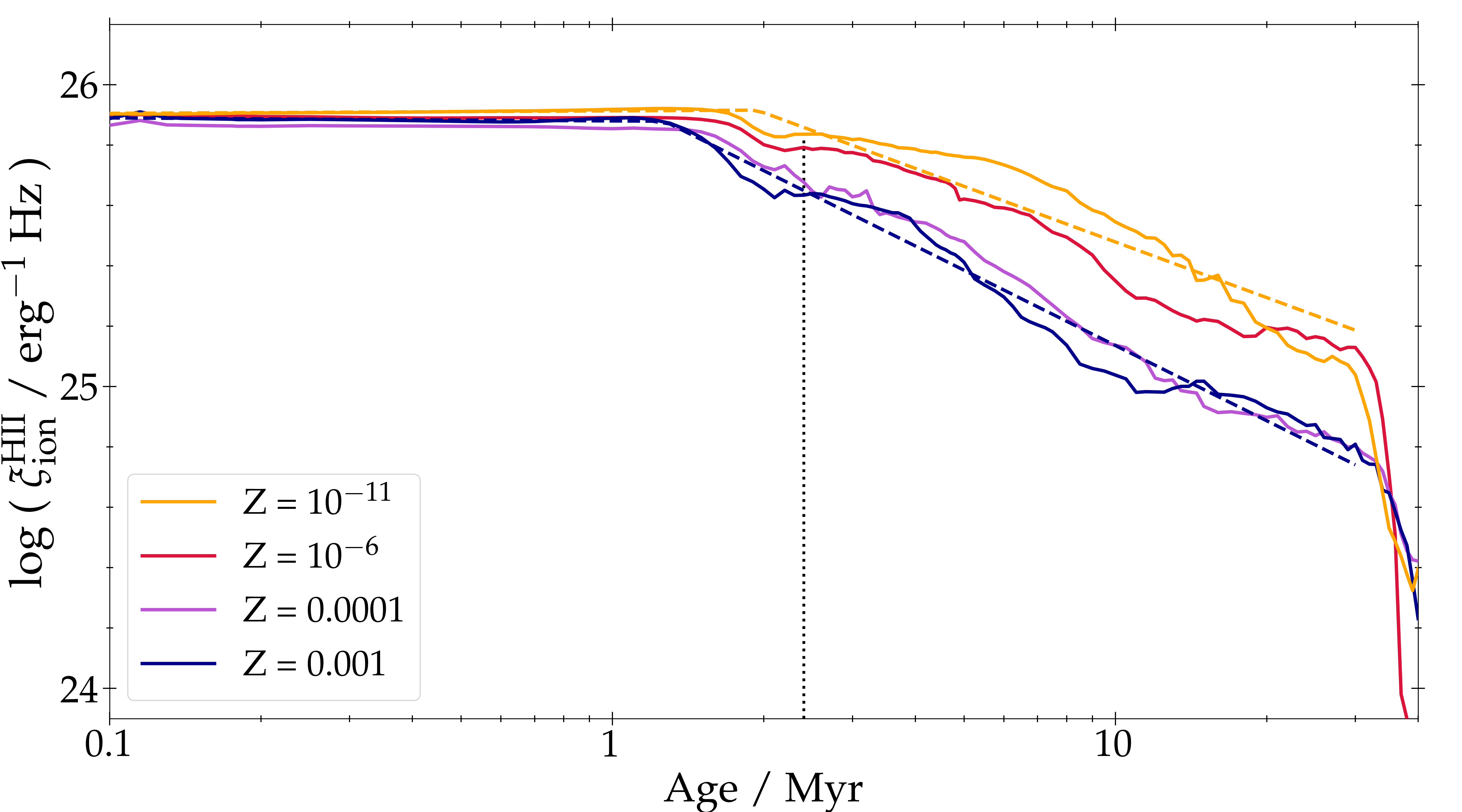}
 \caption{Bimodal linear fits adopted to represent \xiionHII as shown in the previous plot, for the two extreme metallicities $Z = 10^{-11}$ and 0.001. The dashed lines correspond to the adopted linear fits, for the models represented in the same color.}
\label{fig:fit_xiionneb}
\end{figure}

The general shape of the curves in Fig.~\ref{fig:evo_xiionneb} suggests a fit with two linear regimes, before and after the breaks that occur between 1 and 2\,Myr, depending on the metallicity. As the two extreme metallicities, $Z = 10^{-11}$ and 0.001, appear to bracket the intermediate ones, and as the slope after the break appears to decrease monotonically with metallicity, we present fits only for these two metallicities, assuming that the relations for the intermediate metallicities can then be interpolated. We find that
\begin{multline}
\label{eq:fit_xiion_1em11}
 \log\, (\xiionHII / \mathrm{erg}^{-1} \mathrm{Hz}) = \\
     \begin{cases}
        0.0082 \log(\mathrm{t/yr}) + 25.86 & \mathrm{for\,\,\,t} \leq 1.9\,\mathrm{Myr} \\
        -0.61 \log(\mathrm{t/yr}) + 29.77 & \mathrm{for\,\,\,t} > 1.9\,\mathrm{Myr} \\
     \end{cases}
\end{multline}
and
\begin{multline}
\label{eq:fit_xiion_001}
 \log\, (\xiionHII / \mathrm{erg}^{-1} \mathrm{Hz}) = \\
     \begin{cases}
        -0.0094 \log(\mathrm{t/yr}) + 25.93 & \mathrm{for\,\,\,t} \leq 1.1\,\mathrm{Myr} \\
        -0.82 \log(\mathrm{t/yr}) + 30.91 & \mathrm{for\,\,\,t} > 1.1\,\mathrm{Myr} \\
     \end{cases}
\end{multline}
provide reasonable approximations to the actual models for $Z=10^{-11}$ and $Z=0.001$ respectively, as illustrated by Fig.~\ref{fig:fit_xiionneb}.

\begin{figure}          
\centering
\includegraphics[width=\columnwidth]{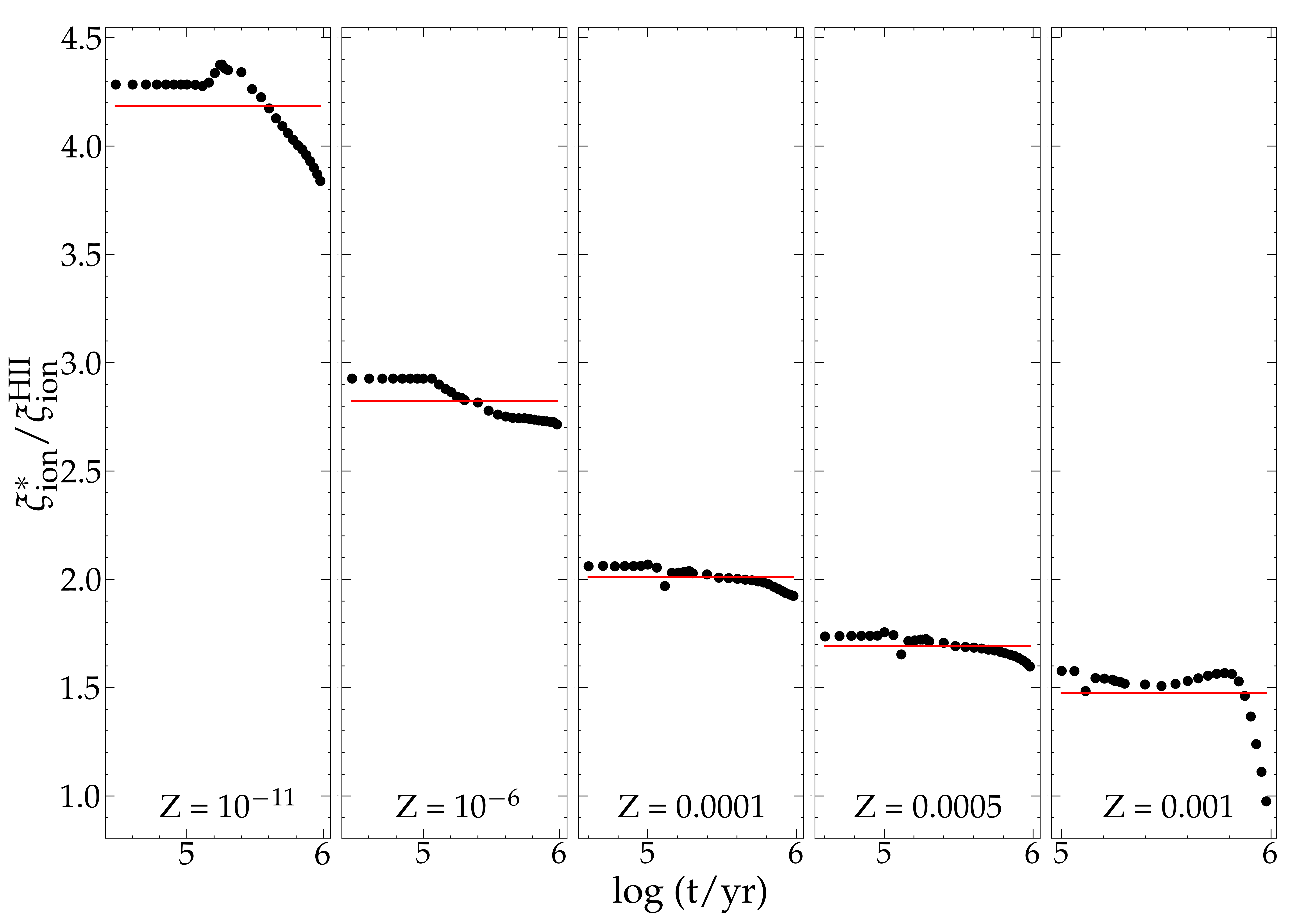}
 \caption{Time evolution of the \xiionst/\xiionHII ratio at ages below 1\,Myr for the same \protect\GLS binary SSP models with a \citet{chabrier2003} IMF as in Fig.~\ref{fig:comp_ndot_VLA}, for different metallicities, shown in different panels. In each panel, the red horizontal line indicates the time-averaged value used to derive the expression of \xiionst/\xiionHII($Z$) in Eq.~\eqref{eq:fit_xiions}.
 }
\label{fig:xiions_ratio}
\end{figure}

\begin{figure}
\centering
\includegraphics[trim=60 30 30 30, clip, width=\columnwidth]{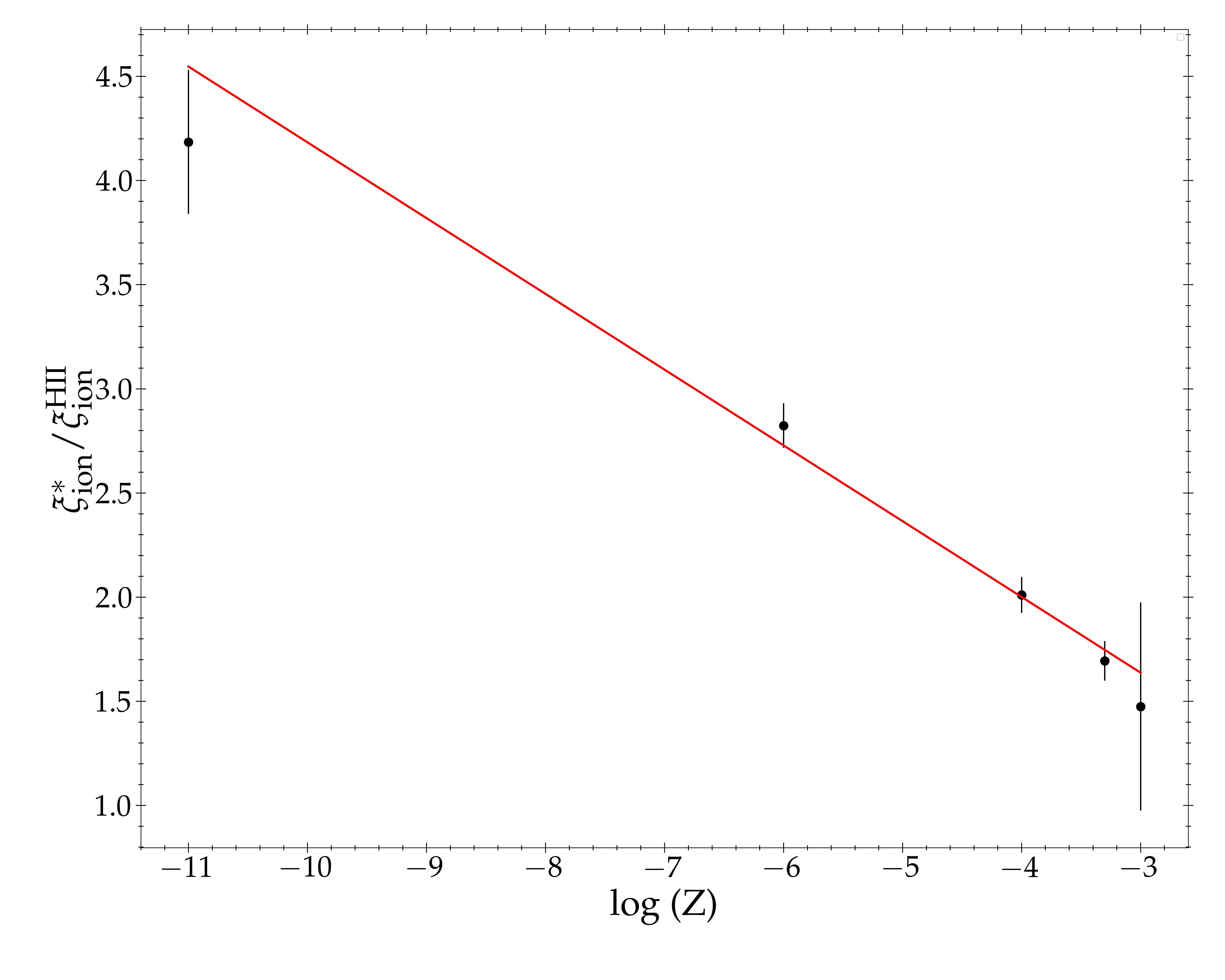}
 \caption{Illustration of the fit to time-average values of \xiionst/\xiionHII at ages of less than 1\,Myr as a function of the metallicity provided by Eq.~\eqref{eq:fit_xiions}. The time-averaged values are estimated from Fig.~\ref{fig:xiions_ratio}. Approximate uncertainties are calculated as the distance between the most extreme point and the mean value at each metallicity (red line) in Fig.~\ref{fig:xiions_ratio}, and are accounted for in the estimation of the best-fitting relation.}
\label{fig:fit_xiions_ratio}
\end{figure}

It is also of interest to provide a means of estimating the stellar ionizing-photon production rate \xiionst, which is often used (in combination with the \fesc parameter) in simulations of the reionization epoch. We can achieve this by combining the above expressions for \xiionHII with an estimate of the ratio between \xiionst and \xiionHII. For simplicity, we focused on the initial, bright phase with nearly constant \xiionHII at ages below 1\,Myr (Fig.~\ref{fig:evo_xiionneb}), during which massive stars evolve on the main sequence. Fig.~\ref{fig:xiions_ratio} shows the evolution of the \xiionst$/$\xiionHII ratio during this phase, for the different metallicities, for the same \protect\GLS binary SSP models with a \citet{chabrier2003} IMF as in Fig.~\ref{fig:comp_ndot_VLA}. The ratio is almost constant during this phase and decreases with increasing metallicity, reflecting the associated increasing absorption of stellar UV luminosity by dust in the \hii region (at fixed dust-to-metal mass ratio \xid). We computed the time-averaged value of this ratio for each metallicity (shown as a red horizontal line in each panel), which we report in Fig.~\ref{fig:fit_xiions_ratio} with uncertainties reflecting the offset of the most extreme value from the time-averaged ratio. The dependence of this time-averaged value on metallicity can be well approximated by the relation 
\begin{equation}
    \xiionst/\xiionHII = -0.35 \log(Z) + 0.67\,,
    \label{eq:fit_xiions}
\end{equation} 
obtained using weights inversely proportional to the uncertainties and shown as the red line in Fig.~\ref{fig:fit_xiions_ratio}.

\subsection{Other properties}
\label{sec:popIII_res_other}

In this subsection, we present complementary quantities predicted by our \protect\GLS models of EoR galaxies, which may offer additional probes of the stellar physics at play in the early Universe. These include the production rate of Lyman-Werner photons and the rates of different types of SNe, which can provide important information about chemical enrichment.

\subsubsection{Lyman-Werner photon production rate}
\label{sec:popIII_res_other_LW}

Unlike most high-energy photons, which are rapidly absorbed by neutral H in the IGM, LW photons have a very long mean free path due to their relatively low energies, just above the Lyman limit. It is therefore important to account for LW photons self-consistently in simulations studying the formation of the first cosmological structures, as this has strong implications for star formation, and hence, for stellar properties, feedback, chemical-enrichment timescales, etc., as explored in recent studies by \citet{gesseyjones2022} and \citet{incatasciato2023}.

\citet{incatasciato2023} present an in-depth study of the LW radiation field, its evolution with redshift and its consequences on H$_2$ photo-dissociation and star formation. They used SPS models to predict LW-photons emission for both Pop\,III and Pop\,II stellar populations. Their calculations were based on the \citet{schaerer2002} models with different IMFs for $Z = 0$, and on \bpass models with different IMFs for Pop\,II ($Z = 0.0005$) populations. \citet{incatasciato2023} conclude that the mean LW intensity must have increased significantly from $z\sim23$ to $z\sim6$ -- primarily due to massive star-forming galaxies -- and highlight the importance of incorporating LW radiation into cosmological simulations for a realistic understanding of early galaxy formation. They examine the complexities of modeling the LW background radiation, comparing various models and noting the impacts of differing assumptions on the predictions. They also investigate the evolving minimum halo mass required for Pop\,III star formation under LW influence, advocating the high sensitivity of such predictions to model limitations. This, and the support by recent JWST observations of actively star-forming galaxies at high redshifts, indicates the crucial role of LW radiation in star formation within low-mass halos in the early Universe.

\begin{figure}
\centering
\includegraphics[width=\columnwidth]{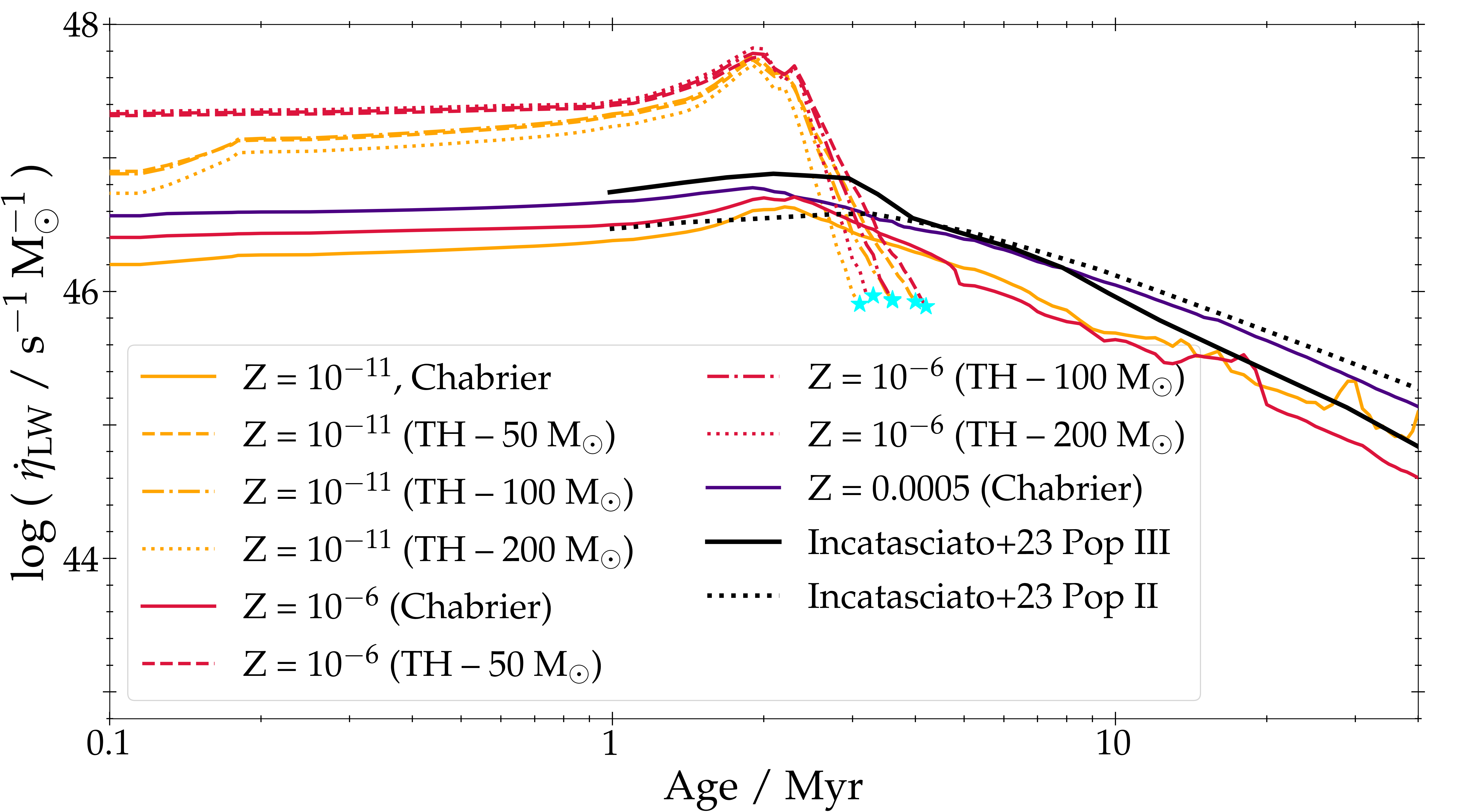}
 \caption{Lyman-Werner photon-production rate (in the wavelength range 912\,\AA\,--\,1150\,\AA) as a function of the stellar-population age for the same \protect\GLS models as in Fig.~\ref{fig:comp_ndot_VLA} (normalized to a total initial stellar mass of 1\,\Msun integrated over 0.1--300\,\Msun). The models with $Z = 0.0001$ and 0.001, which are nearly indistinguishable from the $Z = 0.0005$, are omitted here for clarity. For reference, the predictions of \citet{incatasciato2023} for populations of Pop\,II stars (using $Z = 0.0005$ \bpass models with the \citealt{chabrier2003} IMF) and Pop\,III stars (using zero-metallicity models from \citealt{raiter2010} with a lognormal IMF over 1--500\,\Msun) are shown as dotted and solid black curves, respectively.}
\label{fig:evo_LWrate}
\end{figure}

In Fig.~\ref{fig:evo_LWrate}, we compare the fiducial model of \citet[][available for ages greater than 1\,Myr]{incatasciato2023} for populations of Pop\,II stars (using $Z = 0.0005$ \bpass models with the \citealt{chabrier2003} IMF) and Pop\,III stars (using zero-metallicity models from \citealt{raiter2010} with a lognormal IMF over 1--500\,\Msun) with predictions from the \protect\GLS model. All production rates are normalized to a total initial stellar mass of 1\,\Msun. The production rates of LW photons computed by \citet{incatasciato2023} are in general agreement with those obtained with the \protect\GLS model for a \citet{chabrier2003} IMF. The main difference arises before $\sim3$\,Myr: in the \citet{incatasciato2023} model, Pop\,III stars initially produce more LW photons than their Pop\,II counterparts, before this trend reverses over time. In contrast, the \protect\GLS model predicts that Pop\,II stellar populations consistently produce more LW photons than Pop\,III ones at all times. The \protect\GLS model also reveals a general trend of increasing $\dot{\eta}_{\mathrm{LW}}$ with metallicity at all ages, driven by the gradual shift of the peak wavelength of the hot, near-black-body spectra of the most massive stars on the upper main sequence, from the far-UV to the near-UV. Models with top-heavy IMFs initially exhibit much higher LW photon-production rates (per unit stellar mass), before declining much more steeply shortly after 1\,Myr as the most massive stars die out. This behavior is also observed in the \citet{incatasciato2023} models, as represented in their figure~2.

We find LW photons-production rates in general agreement with those reported by \citet{incatasciato2023}, supporting their conclusions on the importance of including LW radiation in simulations of early galaxy formation and its influence on star formation in low-mass halos. Given the LW photon-production rates predicted here, it is likely that, after the first generations of stars have evolved, most of the molecular H$_2$ in the ISM has been photo-dissociated. At this stage, however, star-forming halos are expected to have reached virial temperatures high enough for atomic cooling to take over, ensuring that star formation continues \citep[e.g.,][]{oh2002}, until metallicity rises enough for metals to become the most efficient coolants.

\subsubsection{Supernova rates}
\label{sec:popIII_res_other_SN}

The study of the different types of SNe occurring in a stellar population provides useful information about the evolutionary paths followed by individual stars. The physical conditions reached during the final core collapse of a star, which are primarily determined by its mass and composition, allow for the differentiation between different types of explosive phenomena. 

Type-Ia SNe arise at the surface of accreting white dwarfs, when their mass exceeds the Chandrasekhar limit for white dwarf stability ($\sim1.4$\,\Msun). These events exhibit characteristic energies on the order of $10^{51}$\,erg, and release relativistic ejecta composed mainly of iron-peak elements due to the runaway fusion reactions in the exploding white dwarf. As indicators of the proportion of low-mass stars in binary systems, type-Ia SNe are also the primary contributors to Fe enrichment in the ISM. On the other hand, type-II SNe are the product of the evolution of massive stars and are responsible for enriching the ISM in $\alpha$ elements (O, Ne, Mg, Si, S, Ca). Pair-instability SNe (PISNe) are particularly interesting as indicators of mass distributions in a stellar population, as they arise only from extremely massive stars, with zero-age MS masses between typically 130 and 250\,\Msun.\footnote{The exact range depends on metallicity, which controls the efficiency of stellar winds and their influence on reaching the He-core mass required to form electron-positron pairs and trigger the PISN mechanism \citep[see, e.g., fig.~2 of][]{spera2017}.} PISNe release enough energy to completely disrupt the stellar remnant, creating a mass gap in the mass spectrum of black holes \citep{spera2015,belczynski2016,woosley2017,spera2017}. Thus, studying PISNe in a stellar population provides insight into how many objects are expected to populate this mass gap. Stars slightly less massive than PISN progenitors, with zero-age MS masses typically between 60 and 130\,\Msun, can still reach He-core masses large enough to convert photons into electron-positron pairs. However, in these cases, the subsequent pair annihilation leads to core contraction and explosive burning of oxygen and silicon, but not enough energy is generated for a complete disruption. Instead, these stars undergo Pulsational PISNe (PPISNe), experiencing multiple pulsations which release significant kinetic energy through mass loss before eventually collapsing into a compact remnant (typically a black hole).

Due to the very high masses of their progenitors and their correspondingly short lifetimes, PISNe and PPISNe are expected to arise only at early ages. Instead, because of the longer evolutionary timescales of low-mass stars and white dwarfs, type-Ia SNe are expected to occur after several tens of millions years. This stark difference in timescales makes the $\alpha$-to-Fe element ratio in the ISM an excellent indicator of the age of a stellar population.

\begin{figure}
\centering
\includegraphics[width=\columnwidth]{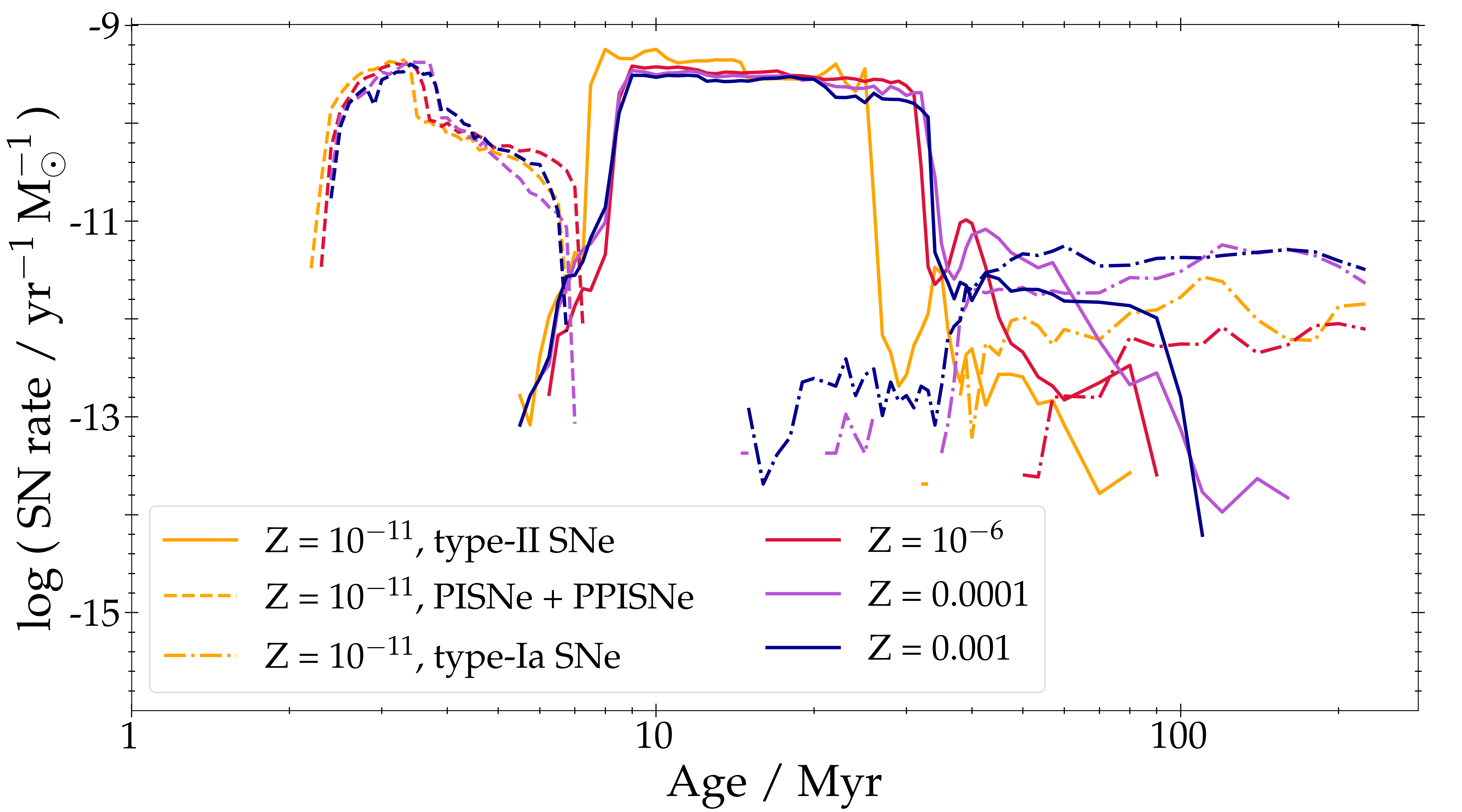}
 \caption{Supernova rate as a function of the stellar-population age for the same binary-star \protect\GLS models with a \citet{chabrier2003} IMF as in Fig.~\ref{fig:comp_ndot_VLA} (normalized to a total initial stellar mass of 1\,\Msun integrated over 0.1--300\,\Msun). Different metallicities are shown in different colors, and different types of SNe are distinguished by line styles (short-dashed for pair-instability, solid for "classical" type-II, and dot-dashed for type-Ia SNe).}
\label{fig:evo_SN}
\end{figure}

Figure~\ref{fig:evo_SN} shows the time evolution of the different types of SNe arising in the same binary-star \protect\GLS SSP models as in Fig.~\ref{fig:comp_ndot_VLA}. We ignore nonexploding, ‘failed’ SNe leading to the formation of black holes \citep{spera2015,spera2017}. For clarity, only models with a \citet{chabrier2003} IMF are represented (predictions for top-heavy IMFs can be inferred from these models using mass-distribution arguments). The five different metallicities are represented in different colors, while different line styles represent the different types of SNe. 

As expected, the PISNe+PPISNe peak in Fig.~\ref{fig:evo_SN} arises between 2 and 6\,Myr across all metallicities. This rate is not strongly affected by metallicity, as the metallicities represented here are all fairly low. For higher metallicities, this rate would decrease, eventually falling to zero for metallicities above $Z = 0.01$ for PISNe and 0.018 for PPISNe \citep{spera2017}.
Other types of type-II SNe, from lower-mass progenitors, emerge around 10\,Myr and persist for a few tens of Myr. The onset and duration of this peak are similar for all metallicities, which is not surprising as the IMF is the same for all models. Yet, the duration of the peak appears slightly shorter at the lowest metallicity, likely due to the faster evolution of the hotter stars at such low metallicity.
Type-Ia SNe first appear around 30\,Myr, displaying a similar behavior across all metallicities.

Finally, we note that chemical enrichment in Pop\,III models begins around 2\,Myr, marked by the appearance of the first PISNe.

\section{Gravitational-wave signal from Pop III binary black holes}
\label{sec:popIII_gw}

As previously mentioned, direct optical observations of Pop\,III stars are extremely difficult. It is therefore useful to focus on indirect observations and multimessenger emission, as these could provide valuable counterparts to the detection of Pop\,III stars at high redshift. Pair-instability and core-collapse SNe, which are expected to reach luminosities of up to $10^{12}$ \Lsun \citep{whalen2013a,whalen2013b}, offer attractive opportunities for indirect detection up to $z\sim15$ \citep{whalen2013a,whalen2013b,whalen2014,smidt2014,smidt2015} and can trigger gamma-ray bursts \citep{wang2012,burlon2016} potentially detectable up to $z\sim20$ with near-future facilities \citep{amati2018}. However, the short duration of such transients makes their detection probability still rather low. Pop\,III stars can also be detected indirectly through the impact of their intense radiation on the timing and depth of the cosmological 21-cm signal. This can be explored via the incorporation of spectral models such as \protect\GLS into simulations of the global 21-cm emission \citep[e.g.,][]{mirocha2018, mebane2020, gesseyjones2022, ventura2023, pochinda2024}. In this section, we focus on another promising indirect probe of the properties of Pop\,III stellar populations: binary black holes (BBHs), and in particular their mergers, which should produce potentially detectable gravitational-wave signatures -- conveniently exempt from foreground contamination and line-of-sight absorption, unlike electromagnetic messengers. In this section, we adopted the most recent Planck values \citep{planck_h0} for all cosmological parameters.\footnote{Specifically, we used the matter and dark-energy density parameters $\Omega_{\mathrm{m}}=0.315$ and $\Omega_{\Lambda}= 0.6847$, and the Hubble constant $H_{0}$~= 67.4\,km\,s$^{-1}$\,Mpc$^{-1}$.}

For the sake of brevity, we do not go into detail about the basics of GW emission by two merging compact objects. We refer the reader to, for example, \citet{maggiore2007,maggiore2024} for thorough reviews on GWs and more details about the gravitational signal emitted by merging BBHs.

\subsection{Binary black-hole mergers in \protect\GLS models of EoR galaxies}
\label{sec:popIII_gw_res}

We followed pure-binary SSPs with a \citet{chabrier2003} IMF for the five metallicities considered in the previous sections. For each BBH system formed, we estimated the merger time based on the BH masses, separation (semi-major axis), and eccentricity at the time of formation, according to the following equation, which provides high accuracy (relative error of less than 0.6\%) across the full range of eccentricities \citep{iorio2023}:
\begin{equation}
    \label{eq:GW_tdelay}
    t_{\mathrm{delay}} =\frac{1}{1 + f_{\mathrm{corr}}(e)} \, \frac{5c^5}{256G^3} \, \frac{a^4}{m_1 m_2 (m_1 + m_2)} (1-e^2)^{7/2}\,,
\end{equation}
with 
\begin{equation}   
f_{\mathrm{corr}}(e) = e^2 \left[ -0.443 + 0.580(1 - e^{3.074})^{1.105 - 0.807 e + 0.193 e^2} \right]
\end{equation}
(see appendix~D of \citealt{iorio2023} for details).


\begin{figure}
\centering
\includegraphics[width=\columnwidth]{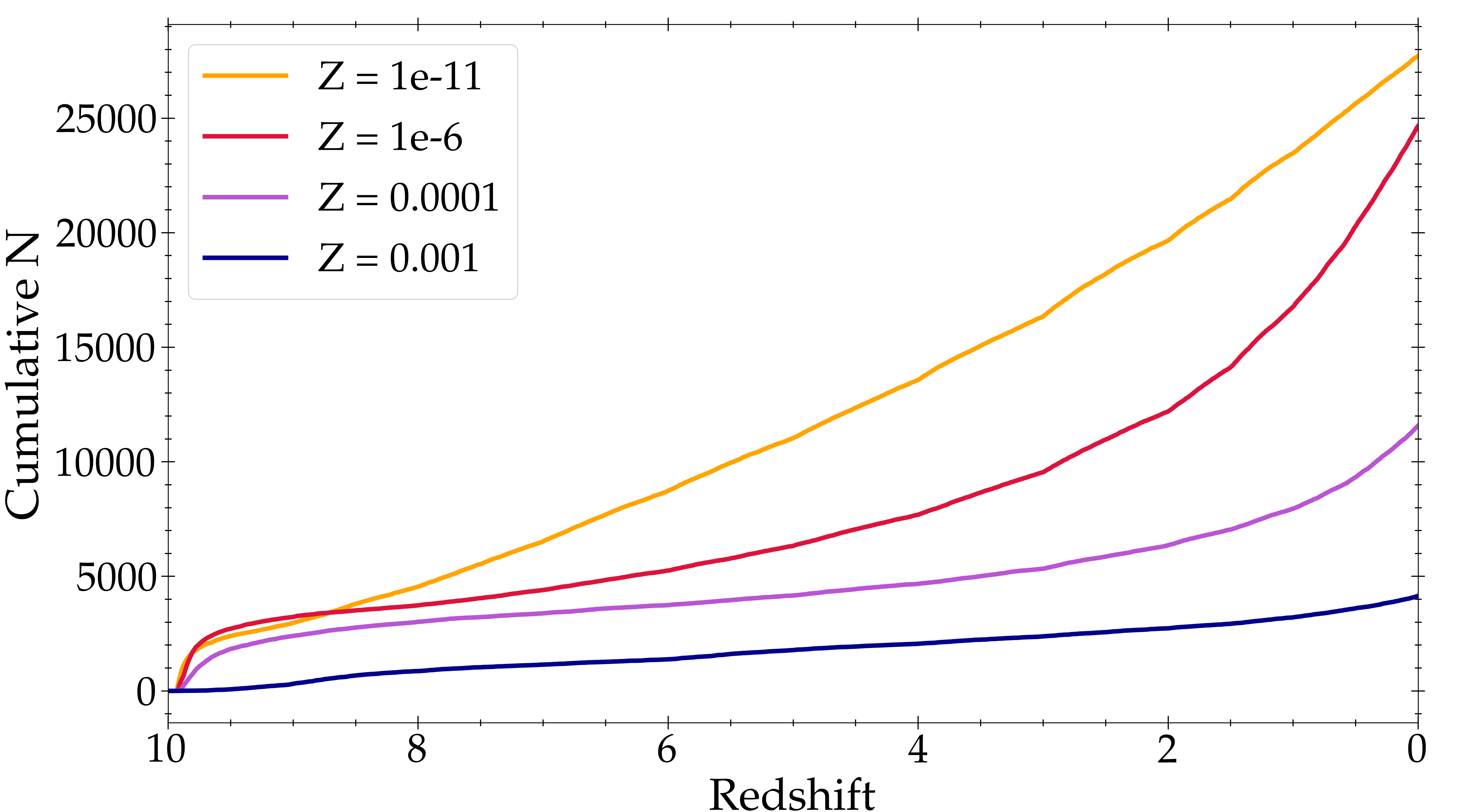}
 \caption{Cumulative number of binary black hole mergers as a function of the redshift in \protect\GLS populations of pure-binary stars (of 1 million pairs each) for different metallicities.}
\label{fig:evo_BBH_mergers}
\end{figure}

The expected cumulative number of merging BBHs in the \protect\GLS models is shown as a function of the redshift in Fig.~\ref{fig:evo_BBH_mergers}. For the Pop\,III model, approximately a third of the total BBH population merges between $z = 10$ and $z = 0$ and contributes to this plot. The number of merging BBHs increases significantly with decreasing metallicity. This results from the less efficient mass loss at such low metallicities, which allows stars to retain more mass, making them larger and less compact, increasing the likelihood of interactions, such as mass-transfer or common-envelope phases, and leading to tighter BBHs able to merge within a Hubble time \citep[e.g.,][]{iorio2023}. 
The number of mergers initially rises rapidly at high redshift, particularly in the most metal-poor models. This is due to the most massive and tightest binaries very rapidly becoming close BBHs and merging, all within a few Myr. Finally, although the total number of merging BBHs is largest for $Z = 10^{-11}$, it is below that for $Z = 10^{-6}$ during the first Myr. This arises from the larger masses of progenitor stars at $Z = 10^{-11}$, which increase the probability of unstable mass transfer resulting in premature mergers and common envelope configurations, hence reducing the formation of very tight BBHs. The surviving, looser BBHs exhibit longer delay times before merging.

\begin{figure}
\centering
\includegraphics[width=\columnwidth]{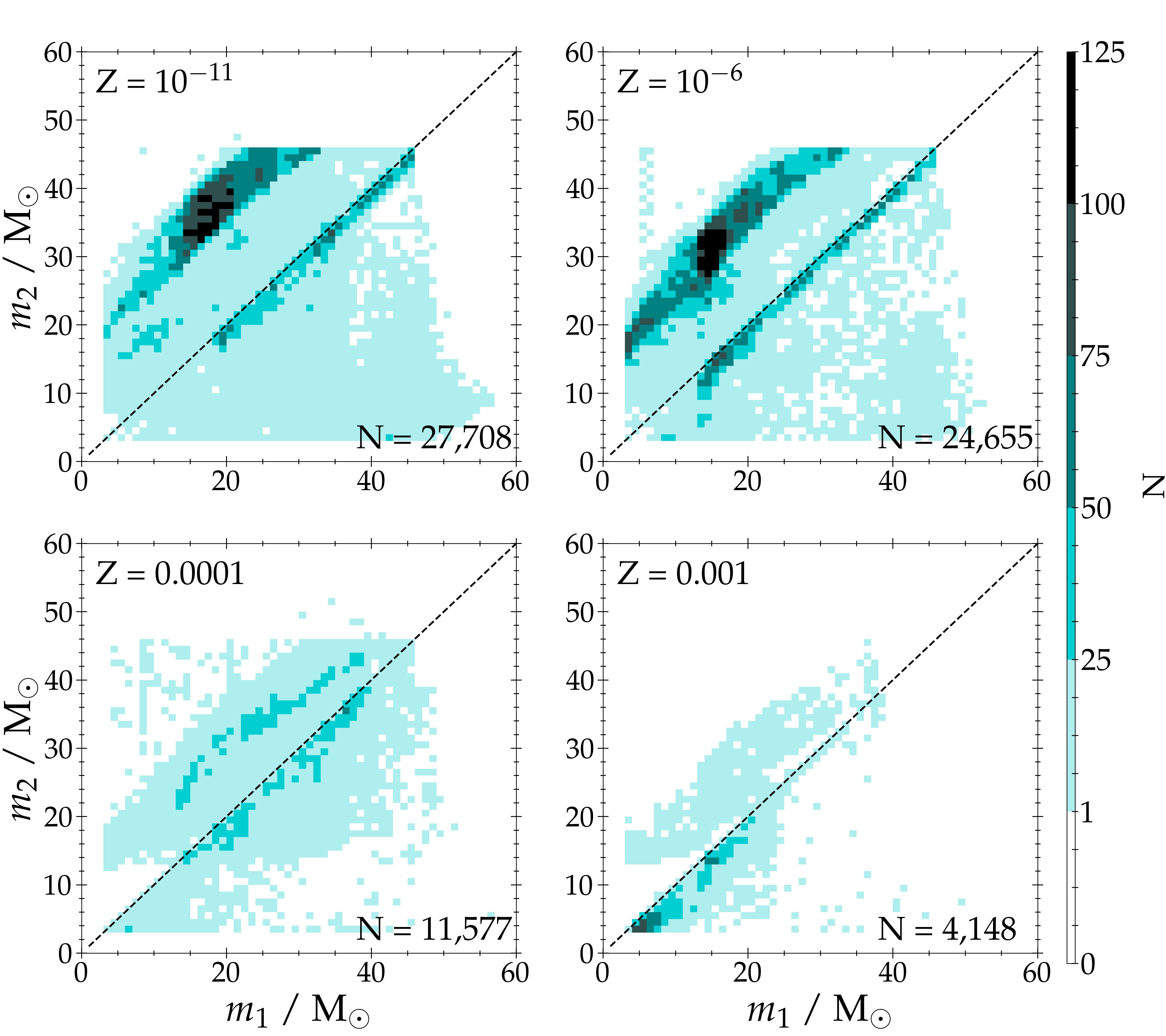}
 \caption{Two-dimensional histograms of the masses of black holes in merging BBH systems, for the same \protect\GLS models as in Fig.~\ref{fig:evo_BBH_mergers}, for the metallicities $Z = 10^{-11}$, $Z = 10^{-6}$, $Z = 0.0001$, and $Z = 0.001$ (in different panels, as indicated). The quantity $m_1$ refers to the mass of the BH formed from the primary star, whereas $m_2$ refers to that of the BH formed from the secondary star. The total number of BBH mergers in the modeled stellar population is indicated in the lower-right corner of each panel. The identity relation is shown as a dashed line.}
\label{fig:BBH_mergers_masses}
\end{figure}

\begin{figure}
\centering
\includegraphics[width=\columnwidth]{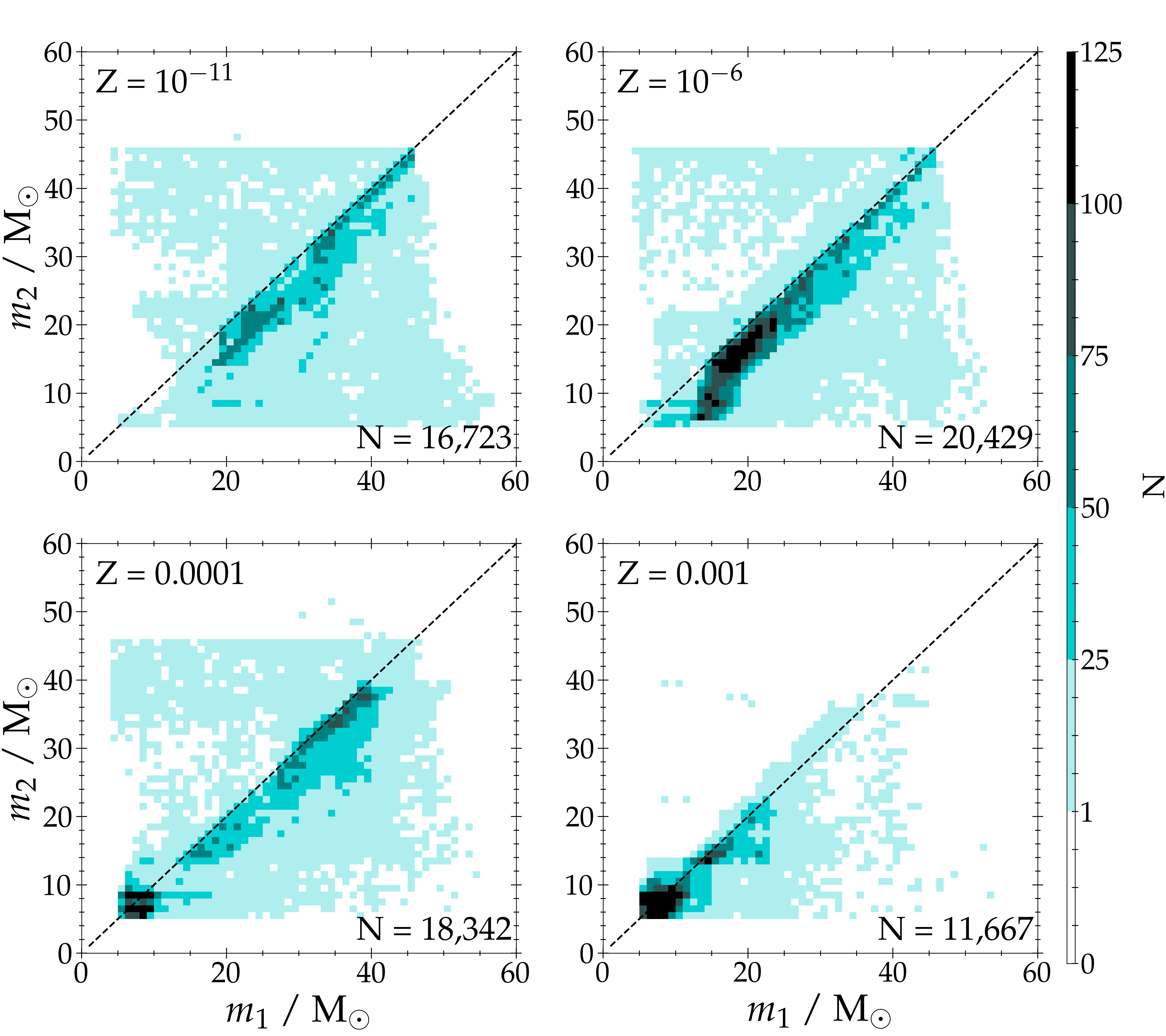}
 \caption{Same as Fig.~\ref{fig:BBH_mergers_masses}, but for binary-star models not including QHE.}
\label{fig:BBH_mergers_masses_QHEeffect}
\end{figure}

We now focus on the masses of BBHs that merge before $z=0$ according to Fig.~\ref{fig:evo_BBH_mergers}. We show this distribution for black holes forming from the primary ($m_1$) and secondary ($m_2$) stars in Fig.~\ref{fig:BBH_mergers_masses}, for the metallicities $Z = 10^{-11}$, $Z = 10^{-6}$, $Z = 0.0001$, and $Z = 0.001$.

Figure~\ref{fig:BBH_mergers_masses} exhibits two notable features: in each panel, the objects tend to be distributed primarily either along the identity relation, or in the upper-left corner of the diagram. This second, increasingly prominent sequence toward low metallicities, reveals that in systems where the black-hole masses are not similar, the remnant of the secondary star tends to be the most massive. 
This is primarily due to the inclusion of quasi-homogeneous evolution in the binary-star models, which results in compact, pure-helium secondary stars immediately after the main sequence following Roche-lobe mass transfer from the primary star. These helium stars lead to the formation of more massive black holes, almost as massive as the pure helium stars, than in models without QHE. To illustrate this, we show in Fig.~\ref{fig:BBH_mergers_masses_QHEeffect} the analog of Fig.~\ref{fig:BBH_mergers_masses}, but for binary-star models not including QHE. The sequence of BBHs with $m_2>m_1$ is largely absent from these diagrams. At the lowest metallicities, the significantly smaller number of merging BBHs in models without QHE is mainly due to premature mergers before the secondary can form a black hole, and partly to greater mass loss from less compact secondaries, giving rise to neutron stars rather than black holes in SNe. At the highest metallicities, instead, the significantly larger number of BBHs merging by $z=0$ in models without QHE results from more efficient mass loss leading to less massive secondaries and tighter systems. Hence, the inclusion of QHE in binary-star models can strongly influence the population of early BBHs that are expected to merge within a Hubble time. 

\subsection{Detectability of binary black-hole mergers from EoR galaxies}
\label{sec:popIII_gw_det}

Having highlighted the variety in orbital and mass parameters of BBHs expected to merge between $z=10$ and now in the \protect\GLS models, it is interesting to investigate what fraction of mergers from Pop\,III precursors could actually be observed by current or near-future GW detectors. We provide in this section an estimate of the percentage of sources from the \protect\GLS populations whose signal-to-noise ratio would be sufficiently high to be detected with the upcoming LIGO-Virgo-KAGRA O5 run and with the Einstein Telescope.

For the Einstein Telescope (ET), we assumed the ET-D configuration, composed of three nested triangular detectors with arms of 10\,km each \citep[e.g.,][]{hild2008,maggiore2020,branchesi2023}. For simplicity, we assumed the three detectors to be identical and independent. 
Furthermore, for simplicity, and given their sensitivities and optimal frequency ranges for detecting merging black holes with masses around 30\,\Msun, we focused on these two experiments only, considering the O5 run for LIGO-Virgo and the predicted specifications for ET \citep[expected first light in 2036,][]{branchesi2023}. We used sensitivity curves from recent publications \citep{abbott2020,maggiore2024}.

For this study, we considered the population of BBHs emerging from a stellar population formed during a single burst at $z = 10$, which would merge before $z = 0$ (using Eq.~\ref{eq:GW_tdelay}, as in the previous section, for this selection).
We made the conventional assumption that a merger would be detectable if it gives a signal-to-noise ratio (S/N) greater than 9 for a given detector. We estimated the S/N of an individual signal following the approach outlined by \citet{santoliquido2023}, taking care of combining all three detectors of ET in the S/N calculation.

For each merging BBH, we evaluated the S/N based on the following equation\footnote{This equation corresponds to the frequency treatment applied in the matched filtering technique, which allows on-the-fly detection of GWs from continuous observations \citep[see, e.g.,][for more details]{dalcanton2014}}:
\begin{equation}
    \label{eq:GW_SNR}
    \mathrm{S/N}^2 = \int_0^{\infty} \mathrm{d}f \frac{4 \lvert \Tilde{h}(f) \rvert^2}{S_n(f)}\,,
\end{equation}
where $S_n(f)$ represents the detector sensitivity curve. The amplitude of the GW in the frequency domain, $\lvert \Tilde{h}(f) \rvert^2$, was computed using the {\scriptsize{PY}\footnotesize{CBC}} library. In line with several recent studies \citep[e.g.,][]{dominik2015,taylor2018,bouffanais2019,chen2021,santoliquido2023}, we adopted the phenomenological waveform {\footnotesize{IMRP}\scriptsize{HENOM}\footnotesize{XAS}} \citep{garciaquiros2020} for compact binary mergers.

To account for the impact of the position of the objects in the sky, we randomly drew the position of each merger, considering isotropically distributed sources. We also randomly sampled the inclination angle $\iota$ of each system. More details on the exact formul\ae{} implemented in this method, in particular for the different wave polarizations, can be found in section~2.5 of \citet{santoliquido2023}. To avoid any bias resulting from the stochastic sampling of these angles over a relatively small number of sources, we computed the number of detectable sources a hundred times for each population, from which we derive a mean value and standard deviation.

For the Pop\,III ($Z = 10^{-11}$) and extremely metal-poor Pop\,II ($Z = 10^{-6}$) models, corresponding to $Z = 10^{-11}$ and $Z = 10^{-6}$, we obtained very similar results. We found that LIGO-Virgo O5 should be able to detect at most a few percent of the sources in the simulated populations -- more precisely, $6.7 \pm 1.9\%$ for $Z = 10^{-11}$, and $10.2 \pm 2.0\%$ for $Z = 10^{-6}$. The BBH mergers that could be detected by the LVK detectors during O5 occur at low redshift ($z \lesssim 1$), and involve mainly systems with BH masses greater than 20\,\Msun. This is in line with the larger fraction of merging BBHs at low redshift shown in Fig.~\ref{fig:evo_BBH_mergers} for $Z = 10^{-6}$, compared to $Z = 10^{-11}$.

On the other hand, ET-D is predicted to be able to detect $90.2 \pm 4.3\%$ of the merging BBHs for $Z = 10^{-11}$, and $90.2 \pm 4.0\%$ for $Z = 10^{-6}$. 
This finding supports the importance of studying the demographics of such predictions, as a substantial amount of observational gravitational-wave constraints may soon be available to put these predictions to the test \citep[e.g.,][]{santoliquido2024}. However, even if most of these objects become detectable in the near future, it will likely remain challenging to conclusively identify detections as signatures of Pop\,III stars.

\section{Summary and conclusions}
\label{sec:popIII_dsc}

In this work, we have used the \protect\GLS model, introduced and discussed in detail by \citet{lecroq2024}, to compute the emission properties of extremely metal-poor stellar populations with $Z\,=\,10^{-11}$ and $Z\,=\,10^{-6}$. We compared these to \protect\GLS models of metal-poor populations with higher metallicities ($Z\,=\,0.0001$, 0.0005, and 0.001), and to the predictions of \citet{nakajima2022}. Our results confirm that the emission-line diagnostics highlighted by \citet{nakajima2022} are effective for identifying ionization driven by zero-age Pop\,III stars, with the \protect\GLS models providing additional information that these criteria can isolate Pop\,III stellar populations at ages $\lesssim~1$\,Myr, as shown by Figs.~\ref{fig:comp_1st-age_zones} and \ref{fig:comp_to_SN}.
We also provided predictions for the production efficiency of ionizing photons by these metal-poor stellar population, useful to investigate their potential role in the reionization of the Universe. To facilitate such analyses, we derived simple analytical expressions linking this efficiency directly to age and metallicity. Furthemore, we presented predictions for the production rate of Lyman-Werner photons, essential to understand the mechanisms of pristine-gas cooling and early star formation, as well as for supernova rates, which provide insight into the timescales of chemical enrichment. Finally, we explored the properties of binary black holes in these \protect\GLS models, including their merger probabilities and the likelihood of detecting these events through gravitational waves. Such detections could serve as critical observational counterparts to the challenging task of directly identifying Pop\,III stars.

The main conclusions of this study can be summarized as follows.

\begin{itemize}

\item We confirm that the ultraviolet and optical H and He emission-line diagnostics introduced by \citet{nakajima2022} (Fig.~\ref{fig:comp_1st-age_zones}) are effective to discriminate between ionization by very young ($\lesssim~1$\,Myr) Pop\,III stellar populations and other primordial or later sources (Fig.~\ref{fig:comp_to_SN}). We also confirm, as already shown in previous studies \citep[e.g.,][]{eldridge2012,lecroq2024}, that the production of ionizing photons beyond the first million years is dominated by processes originating from binary interactions (Figs.~\ref{fig:comp_ndot_VLA}--\ref{fig:comp_ndot_schaerer}).

\item We find that \protect\GLS models of Pop\,III and metal-poor Pop\,II stellar populations predict high production efficiencies of ionizing photons. Specifically, we find elevated values of \xiionHII relative to standard predictions, consistent with recent observational studies of galaxies in a wide range of redshifts (Fig.~\ref{fig:evo_xiionneb}). We also provide analytical expressions to describe the evolution of \xiionHII as a function of the age and metallicity, and to relate \xiionHII with \xiionst, the latter being commonly used in simulations (Figs.~\ref{fig:fit_xiionneb}--\ref{fig:fit_xiions_ratio}). Comparison of our predictions with recent theoretical and observational results suggests that the escape of only a few percent of ionizing photons from nascent galaxies with such properties would be sufficient to reionize the primordial Universe.

\item We find production rates of Lyman-Werner photon in agreement with recent simulations \citep[e.g.,][]{incatasciato2023}, highlighting the critical role of these photons in the mechanisms and timescales of early star formation (Fig.~\ref{fig:evo_LWrate}).

\item Our analysis of pair-instability, type-II, and type-Ia supernova rates reveals relatively consistent behavior across metallicities ranging from $10^{-11}$ to 0.001. This analysis allows us to identify the onset of chemical enrichment in \protect\GLS Pop\,III stellar populations at approximately 2\,Myr.

\item By examining the properties of binary black holes originating from \protect\GLS Pop\,III stellar populations, we find that the predicted merger rates and mass distributions of these systems depend critically on the inclusion of quasi-homogeneous evolution in the models. Our results suggest that most BBH mergers from such Pop\,III stellar populations, even at high redshifts, should be detectable with the Einstein Telescope. This high detectability positions BBH mergers as an important complement to direct observations in characterizing Pop\,III stellar populations in the EoR.
    
\end{itemize}


\begin{acknowledgements}
We thank the referee for useful comments.
We are grateful to D.~Schaerer and K.~Nakajima for sharing electronic versions of their models and to M. Dall'Amico for helpful discussions.
G.I. acknowledges financial support under the National Recovery
and Resilience Plan (NRRP), Mission 4, Component 2, Investment 1.4, - Call for tender No. 3138 of 18/12/2021 of Italian Ministry of University and Research funded by the European Union – NextGenerationEU.
F.S. acknowledges financial support from the AHEAD2020 project (grant agreement n. 871158).
M.M. acknowledges financial support from the German Excellence Strategy via the Heidelberg Cluster of Excellence (EXC 2181 - 390900948) STRUCTURES. M.M., F.S., G.C., and G.I. also acknowledge financial support from the European Research Council for the ERC Consolidator grant DEMOBLACK, under contract No. 770017 (PI: Mapelli).
G.B. acknowledges financial support from the National Autonomous University of M\'exico (UNAM) through grants DGAPA/PAPIIT IG100319 and BG100622.
\end{acknowledgements}


\bibliographystyle{aa}
\bibliography{paper} 

\end{document}